\documentclass[preprintnumbers,superscriptaddress,amsmath,amssymb,nofootinbib,twocolumn,showpacs]{revtex4}

\usepackage{dcolumn}
\usepackage{bm}
\usepackage{graphicx}

\usepackage[normalem]{ulem}

\usepackage{footmisc}

\usepackage{color}
\usepackage{natbib}
\usepackage{twoopt}

\newcommand{\beq}{\begin{equation}}
\newcommand{\eeq}{\end{equation}}
\newcommand{\bsubeq}{\begin{subequations}}
\newcommand{\esubeq}{\end{subequations}}
\newcommand{\ben}{\begin{eqnarray}}
\newcommand{\een}{\end{eqnarray}}
\newcommand{\balignt}{\begin{alignat}}
\newcommand{\ealignt}{\end{alignat}}
\newcommand{\bi}{\begin{itemize}}
\newcommand{\ei}{\end{itemize}}
\newcommand{\nn}{\nonumber}

\newcommand{\ie}{\mbox{\it i.e.}}
\newcommand{\eg}{\mbox{\it e.g.}}

\newcommand{\citeeq}[1]{Eq.~(\ref{#1})}

\newcommand{\citesec}[1]{Sect.~\ref{#1}}

\newcommand{\citeapp}[1]{App.~\ref{#1}}

\newcommand{\citetab}[1]{Tab.~\ref{#1}}
\newcommand{\citefig}[1]{Fig.~\ref{#1}}

\newcommand{\rhoc}{{\ifmmode {\rho_{\rm c}} \else $\rho_{\rm c}$\fi}}
\newcommand{\rhos}{{\ifmmode {\rho_{\rm s}} \else $\rho_{\rm s}$\fi}}
\newcommand{\rhotilde}{{\ifmmode \tilde{\rho} \else $\tilde{\rho}$\fi}}
\newcommand{\rs}{{\ifmmode {r_{\rm s}} \else $r_{\rm s}$\fi}}
\newcommand{\rt}{{\ifmmode {r_{\rm t}} \else $r_{\rm t}$\fi}}

\newcommand{\Rsun}{\mbox{$R_{\odot}$}}

\newcommand{\mchi}{{\ifmmode m_\chi \else $m_\chi$\fi}}
\newcommand{\sigv}{{\ifmmode \langle \sigma v \rangle \else $\langle \sigma v \rangle$\fi}}

\newcommand{\angleave}[2]{\ifmmode \left \langle{#1} \right \rangle_{#2} \else $\left\langle{#1}\right\rangle_{#2}$ \fi}

\newcommand{\gevpcc}{{\ifmmode {\rm GeV/cm^{3}} \else ${\rm GeV/cm^{3}}$\fi}}

\newcommand{\lcdm}{{\ifmmode \Lambda{\rm CDM} \else $\Lambda{\rm CDM}$\fi}}

\def\usine{\textsc{usine}}

\def\ams{\textsc{Ams-02}}
\def\calet{\textsc{Calet}}
\def\dampe{\textsc{Dampe}}
\def\fermi{\textsc{Fermi}}
\def\pamela{\textsc{Pamela}}
\def\voyager{\textsc{Voyager}}

\def\big{\textsf{BIG}}
\def\slim{\textsf{SLIM}}
\def\quaint{\textsf{QUAINT}}

\def\BIG{\textsf{BIG}}
\def\SLIM{\textsf{SLIM}}
\def\QUAINT{\textsf{QUAINT}}

\def\min{MIN}
\def\med{MED}
\def\max{MAX}

\def\ppc4dm{\texttt{PPC4DM}}

\def\hesse{\textsc{hesse}}
\def\mcmc{\textsc{mcmc}}
\def\migrad{\textsc{migrad}}
\def\minuit{\textsc{minuit}}
\def\minos{\textsc{minos}}
\def\pybind{\texttt{PyBind11}}
\def\pymc{\texttt{PyMC3}}

\def\aap{A\&A}%
\def\apj{ApJ}%
\def\apjl{ApJ}%
\def\apss{Ap\&SS}%
\def\araa{ARA\&A}%
\def\epjc{Europ.~Phys.~J.~C}%
\def\jcap{J. Cosmology Astropart. Phys.}%
\def\jgr{J.~Geophys.~Res.}%
\def\mnras{MNRAS}%
\def\nat{Nature}%
\def\physrep{Phys.~Rep.}%
\def\physscr{Phys.~Scr}%
\def\prd{Phys.~Rev.~D}%
\def\prl{Phys.~Rev.~Lett.}%
\def\ssr{Space~Sci.~Rev.}%

\usepackage[pdftex,
  bookmarks=false,
  breaklinks=true,%
  colorlinks=true,%
  linkcolor=red,
  urlcolor=blue,
  citecolor=blue,
  pdfauthor={Y.~Genolini et al.},%
  pdftitle={New minimal, median, and maximal propagation models for dark matter searches with Galactic cosmic rays}%
]{hyperref}

\DeclareGraphicsExtensions{.png,.pdf}

\graphicspath{{./}{Figs/}}

\begin{document}

\title{New minimal, median, and maximal propagation models\\ for dark matter searches with Galactic cosmic rays}

\author{Yoann G\'enolini}
\email{Yoann.Genolini@nbi.ku.dk}
\affiliation{Niels Bohr International Academy \& Discovery Center, Niels Bohr Institute,
  University of Copenhagen, Blegdamsvej 17, DK-2100 Copenhagen --- Denmark}

\author{Mathieu Boudaud}
\email{Deceased}
\affiliation{Instituto de F\'isica Te\'orica UAM/CSIC, Calle Nicol\'as Cabrera 13-15, Cantoblanco E-28049 Madrid --- Spain}

\author{Marco Cirelli}
\email{marco.cirelli@gmail.com}
\affiliation{LPTHE,
   CNRS \& Sorbonne University, 4 place Jussieu 75252 Paris CEDEX 05 --- France}

\author{Laurent Derome}
\email{laurent.derome@lpsc.in2p3.fr}
\affiliation{LPSC, Universit\'e Grenoble Alpes, CNRS/IN2P3, 53 avenue des Martyrs,
  38026 Grenoble --- France}

\author{Julien Lavalle}
\email{lavalle@in2p3.fr}
\affiliation{Laboratoire Univers \& Particules de Montpellier (LUPM),
  CNRS \& Universit\'e de Montpellier (UMR-5299),
  Place Eug\`ene Bataillon,
  F-34095 Montpellier Cedex 05 --- France}

\author{David Maurin}
\email{david.maurin@lpsc.in2p3.fr}
\affiliation{LPSC, Universit\'e Grenoble Alpes, CNRS/IN2P3, 53 avenue des Martyrs, 38026
  Grenoble --- France}

\author{Pierre Salati}
\email{pierre.salati@lapth.cnrs.fr}
\affiliation{LAPTh,
  Universit\'e Savoie Mont Blanc \& CNRS,
  Chemin de Bellevue,
  74941 Annecy Cedex --- France}

\author{Nathanael Weinrich}
\affiliation{LPSC, Universit\'e Grenoble Alpes, CNRS/IN2P3, 53 avenue des Martyrs, 38026
  Grenoble --- France}

\begin{abstract}
Galactic charged cosmic rays (notably electrons, positrons, antiprotons and light antinuclei) are powerful probes of dark matter annihilation or decay, in particular for candidates heavier than a few MeV or tiny evaporating primordial black holes. Recent measurements by \pamela{}, \ams{}, or \voyager{} on positrons and antiprotons already translate into constraints on several models over a large mass range. However, these constraints depend on Galactic transport models, in particular the diffusive halo size, subject to theoretical and statistical uncertainties. We update the so-called \min-\med-\max{} benchmark transport parameters that yield generic minimal, median and maximal dark-matter induced fluxes; this reduces the uncertainties on fluxes by a factor of about 2 for positrons and 6 for antiprotons, with respect to their former version. We also provide handy fitting formulae for the associated predicted secondary antiproton and positron background fluxes. Finally, for more refined analyses, we provide the full details of the model parameters and covariance matrices of uncertainties.

\end{abstract}

\maketitle
\preprint{LAPTH-011/21, LUPM:21-002}
%
%
%

%
\section{Introduction}
\label{sec:intro}
The nature of the dark matter (DM) that dominates the matter content of the Universe is still mysterious. DM is at the basis of our current understanding of structure formation, and it is one of the pillars of the standard $\Lambda$-cold DM ($\Lambda$CDM) cosmological scenario \cite{Peebles1982,BlumenthalEtAl1984,BertoneEtAl2018}, despite potential issues on sub-galactic scales (\eg~\cite{BullockEtAl2017}).

The recent decades have seen a number of different strategies put in place in order to find explicit manifestations of DM, which to date has been detected only gravitationally. For instance, indirect searches (see \eg~\cite{Bergstroem2000,LavalleEtAl2012,Gaskins2016}) aim at discovering potential signals of annihilations (or decays) of DM particles from galactic to cosmological scales, in excess to those produced by conventional astrophysical processes. In this paper, we focus on searches for traces of DM annihilation or decay in the form of high-energy antimatter cosmic rays (CRs) \cite{SilkEtAl1984,RudazEtAl1988,JungmanEtAl1996,DonatoEtAl2000,Bergstroem2000,MaurinEtAl2002,PorterEtAl2011,CirelliEtAl2011,LavalleEtAl2012,AramakiEtAl2016}, the latter being barely produced in conventional astrophysical processes. Prototypical DM candidates leading to this kind of signals are weakly-interacting massive particles (WIMPs), which are currently under close experimental scrutiny \cite{ArcadiEtAl2017,LeaneEtAl2018}. WIMPs are an example of thermally-produced particle DM in the early Universe, a viable and rather minimal scenario \cite{ZeldovichEtAl1975,LeeEtAl1977a,GunnEtAl1978,DolgovEtAl1981,BinetruyEtAl1984a,SrednickiEtAl1988,SteigmanEtAl2012}, in which effective interactions with visible matter may be similar in strength to the standard weak one (see \eg~\cite{Feng2010}). In this paper, for illustrative purpose, we restrict ourselves to the conventional case, i.e. annihilations that produce as much matter as antimatter with speed-independent cross sections ($s$-wave). However, we stress that the benchmark propagation models we derive below are relevant for a variety of other exotic processes or candidates, for which antimatter is also a powerful probe (\eg~decaying DM, $p$-wave processes \cite{BoudaudEtAl2019a}, complex exotic dark sectors \cite{AlexanderEtAl2016}, or evaporation \cite{Hawking1974} of very light primordial black holes \cite{Turner1982,MacGibbonEtAl1991,BarrauEtAl2002,BoudaudEtAl2019}).

The source term composition, spectrum, and spatial distribution (for DM-produced CRs) are dictated by both the fundamental properties of particle DM (mass, annihilation channels, annihilation rates), and its distribution properties on macroscopic scales. Once injected, the produced CRs propagate in the turbulent magnetized Galactic environment where they experience diffusion-loss processes, which modify their initial features. Those reaching Earth constitute the potential DM signal in positrons and antiprotons (\eg~\cite{ReinertEtAl2018}), for many space-borne experiments; notably \pamela{}~\cite{PicozzaEtAl2007,AdrianiEtAl2014a}, \fermi{}~\cite{AtwoodEtAl2009}, \ams{}~\cite{BattistonEtAl2008,Kounine2012,AguilarEtAl2013}, \calet{}~\cite{BrogiEtAl2020}, \dampe{}~\cite{ChangEtAl2017}, and \voyager{}~\cite{StoneEtAl1977,StoneEtAl2013}\footnote{Although launched more than 40 years ago, the latter missions are now playing a decisive role in the understanding of CR propagation \cite{CummingsEtAl2016} as well as in constraining DM \cite{BoudaudEtAl2017,BoudaudEtAl2019}.}.

A key aspect in making predictions is therefore to assess, as realistically as possible, the systematic or theoretical uncertainties affecting Galactic CR transport (\eg~\cite{StrongEtAl2007,GrenierEtAl2015,AmatoEtAl2018,Serpico2018,GabiciEtAl2019,EvoliEtAl2019,DeromeEtAl2019}).
An effective way of providing the signal theoretical uncertainties was proposed in \cite{DonatoEtAl2004}. In this reference, these authors introduced three benchmark sets of Galactic propagation parameters (known as \min, \med, and \max), picked among those best-fitting B/C data at that time \cite{MaurinEtAl2001}. Of these, \med\ was actually the best-fit model, while \min\ and \max\ were respectively minimizing  and maximizing the antiproton signal predictions in some specific supersymmetric DM scenarios. These models were also later used for predictions of the positron signal, together with some dedicated analogues~\cite{DelahayeEtAl2008}. They have been exploited by a broad community for mainly two reasons: first, the overall propagation framework was minimal enough to allow for simple analytic or semi-analytic solutions to the CR transport equation \cite{GinzburgEtAl1964,BerezinskyEtAl1990,PtuskinEtAl1996,JonesEtAl2001,MaurinEtAl2001,MaurinEtAl2002,TailletEtAl2004a,MaurinEtAl2006,Maurin2020}; second, the associated astrophysical backgrounds were generally provided in dedicated studies, so that setting constraints on DM candidates only required calculations of the exotic signal. These models were eventually included in public tools for DM searches, such as \textsc{Micromegas}~\cite{BelangerEtAl2011} or \textsc{PPPC4DMID}~\cite{CirelliEtAl2011}. However, as a consequence of the plethora of increasingly precise CR data, they are now outdated (see \eg~\cite{LavalleEtAl2014,BoudaudEtAl2017a,GenoliniEtAl2019,WeinrichEtAl2020}).

In this paper, we propose new benchmark \min-\med-\max\ parameters for the three transport schemes introduced in \cite{GenoliniEtAl2019}\footnote{These schemes (denoted \big, \slim, and \quaint) have slightly different parametrizations of the diffusion coefficient and include or not reacceleration. See \cite{GenoliniEtAl2019} or \citeapp{app:1D2D} (in this paper) for details.}. These parameters are based on the analysis of the latest {\sc Ams-02} \cite{AguilarEtAl2016a,AguilarEtAl2017,AguilarEtAl2018,AguilarEtAl2018b,AguilarEtAl2019b} secondary-to-primary ratios\footnote{Secondary CRs are species absent from sources and only created by nuclear fragmentation of heavier species during their transport, while primary species are CRs present in sources and sub-dominantly created during transport. Typical secondary-to-primary ratios are $^3$He/$^4$He, Li/C, Be/C, and B/C.} \cite{WeinrichEtAl2020}, and fully account for the breaks observed in the diffusion coefficient at high \cite{AhnEtAl2009,AhnEtAl2010,Tomassetti2012,BlasiEtAl2012a,AguilarEtAl2015,GenoliniEtAl2017,Blasi2017,GenoliniEtAl2019} and low rigidity \cite{GenoliniEtAl2019,VittinoEtAl2019}; these parameters also provide secondary antiprotons fluxes consistent \cite{BoudaudEtAl2020} with {\sc Ams-02} data \cite{AguilarEtAl2016}. In contrast to the previous \min-\med-\max\ benchmarks, we now account for the constraints on the diffusive halo size $L$ derived in \cite{WeinrichEtAl2020b}; these constraints are set by radioactive CR species \cite{DonatoEtAl2002} and to a lesser extent by the no-overshoot condition on secondary positrons \cite{LavalleEtAl2014}. A further improvement is that we devise a sound statistical method to pick (in the hyper-volume of allowed propagation parameters) more representative benchmarks, so that our new \min-\med-\max\ sets are now valid for both antiprotons (and more generally, for light antinuclei) {\it and} positrons.

Computing the minimal, median, and maximal exotic fluxes with our new benchmarks allows to easily go from conservative to aggressive new physics predictions, i.e.~to bracket the uncertainty on the detectability of DM candidates. However, in principle, the \min-\med-\max\ sets are more suited to derive constraints than to seek excesses in antimatter CR data. Indeed, the latter should rely on full statistical analyses including correlations of errors (\eg~\cite{Winkler2017,BoudaudEtAl2020,ReinertEtAl2018,HeisigEtAl2020}). For more elaborate comparisons with existing data, one can still perform the full CR analysis with \usine~\cite{Maurin2020,DeromeEtAl2019}\footnote{\href{https://dmaurin.gitlab.io/USINE}{https://dmaurin.gitlab.io/USINE}}, or with other complementary codes like \textsc{Galprop}~\cite{StrongEtAl1998}\footnote{\href{https://galprop.stanford.edu/}{https://galprop.stanford.edu/}}, \textsc{Dragon}~\cite{EvoliEtAl2008,EvoliEtAl2017}\footnote{{\href{https://github.com/cosmicrays/}{ https://github.com/cosmicrays/}}}, or \textsc{Picard}~\cite{Kissmann2014,KissmannEtAl2015}\footnote{\href{https://astro-staff.uibk.ac.at/~kissmrbu/Picard.html}{https://astro-staff.uibk.ac.at/~kissmrbu/Picard.html}}.

The paper develops as follows. In \citesec{sec:generalities}, we quickly review the formalism of Galactic CR transport and the basic ingredients entering the DM source term. In \citesec{sec:primaries}, we pedagogically derive parameter combinations that drive the DM-produced antimatter signals, for antiprotons and positrons in turn. In \citesec{sec:stats}, we introduce the statistical method with which we define the sets of CR transport parameters that maximize and minimize the exotic CR fluxes. We present our results in \citesec{sec:results}, where we also show comparisons between fluxes calculated from the old and new benchmarks. We conclude in \citesec{sec:concl}.

We postpone to appendices some practical pieces of information and more detailed considerations. In particular, App.~\ref{app:1D2D} may prove useful for many readers, as it gathers transport parameters best-fit values and their covariance matrices of uncertainties; the latter can be used for a more evolved DM analysis, going beyond the simple use of \min-\med-\max{}. We also provide parametric formulae (and ancillary files) for the astrophysical background prediction (both antiprotons and positrons). In App.~\ref{app:mcmc}, we perform a cross-validation of the uncertainties and correlations on the transport parameter. Whereas the main text focuses on the \slim{} propagation scheme, App.~\ref{app:min_med_max_general} extends the discussion to the \big{} and \quaint{} schemes.

\section{Generalities}
\label{sec:generalities}
\subsection{Galactic cosmic-ray transport}
In this section, we shortly recall the formalism and the main ingredients of Galactic CR transport and of the DM source term. These will be instrumental to discuss, in the following section, the scaling of the DM-produced antimatter fluxes with the main propagation parameters, in order to motivate the way we will define our \min-\med-\max\ configurations. 

The generic steady-state diffusion-loss equation for a CR species $a$ in energy ($E$) space (differential density per unit energy $\psi_a\equiv dn_a/dE$) is \cite{GinzburgEtAl1964}
\ben
\label{eq:prop}
&-& \vec{\nabla}_{\bf x} \left( K\vec{\nabla}_{\bf x}\psi_a -
\vec{V}_{\rm c} \psi_a \right) 
- \frac{\partial}{\partial E} \left( b_{\rm loss}\psi_a + \beta^2 K_{pp}\frac{\partial\psi_a}{\partial E} \right)\nonumber\\
&&= {\cal Q}_a^{\rm src\;prim}+\sum_b \Gamma_{b\to a}^{\rm src\;sec} \; \psi_b
- \Gamma_a^{\rm sink} \; \psi_a\;.
\een
The first line describes the spatial diffusion $K(E)$ and convection $V_{\rm c}$, and the energy transport with energy losses $b_{\rm loss}(E)\equiv-dE/dt$ and momentum diffusion $K_{pp}$. More details are given in \citeapp{app:1D2D}. In particular, the complete forms of the diffusion coefficients used in this work are given in \citeeq{eq:def_K} (diffusion in real space) and in \citeeq{eq:def_Kpp} (diffusion in momentum space). The second line corresponds to the source and sink terms that are listed below.
\begin{itemize}
   \item The source terms include a primary contribution ${\cal Q}_a^{\rm src\;prim}$, and secondary contributions $\Gamma_{b\to a}^{\rm src\;sec} \equiv \sigma_{ba}\, v_b \, n_{\rm ism} +{\cal BR}_{ba}/(\gamma_b\tau_b$). The latter arises from the sum over (i) inelastic processes converting heavier species of index $b$ into $a$ species (production cross section $\sigma_{ba}$ from impinging CR at velocity $v_b$ on the InterStellar Medium density $n_{\rm ism}$ at rest), and (ii) decay of unstable species with a decay rate $\tau_b$ and branching ratio ${\cal BR}_{ba}$ ($\gamma_b$ is the Lorentz factor of species $b$). Note that for DM products, these secondary contributions are part of the conventional astrophysical background (there are very likely also conventional astrophysical sources of primary positrons \cite{AharonianEtAl1995,HooperEtAl2009,DelahayeEtAl2010} and antiprotons \cite{BlasiEtAl2009}, which should add up to the background).
   \item The sink terms, $\Gamma_a^{\rm sink} \equiv \sigma_a \, v_a \, n_{\rm ism} + 1/(\gamma_a\tau_a)$, include inelastic interactions on the ISM (destruction of $a$) and decay---these terms are irrelevant for positrons.
\end{itemize}

In this study, the transport equation is solved semi-analytically in a magnetic slab of half-height $L$ (and radial extent $R$) that encompasses the disk of the Galaxy, and inside which, the spatial diffusion coefficient is assumed to be homogeneous. The derived interstellar (IS) flux predictions are then compared to top-of-atmosphere (TOA) data by means of the force-field approximation \cite{GleesonEtAl1968a,Fisk1971}, which allows to account for solar modulation effects \cite{Potgieter2013}. The modulation level appropriate to any dataset can be inferred from neutron monitor data \cite{GhelfiEtAl2017}, and be retrieved online from the CR database (CRDB)\footnote{\url{https://lpsc.in2p3.fr/crdb}, see `Solar modulation' tab.} \cite{MaurinEtAl2014,MaurinEtAl2020}.

All transport processes introduced above are characterized by free parameters of a priori unknown magnitude, except for energy loss, inelastic scattering, or decay which depend on constrained ingredients. These free parameters are usually fitted to CR data, in particular to the secondary-to-primary CR ratios (the most conventionally used being the boron-to-carbon (B/C) ratio), in which the source term almost cancels out. Using AMS-02 B/C data, three different propagation schemes, motivated by microphysical considerations, were introduced in \cite{GenoliniEtAl2019}; these schemes are detailed in \citeapp{app:1D2D}. For each of them, we use the associated propagation parameters constrained by the recent analysis of AMS-02 Li/C, Be/C, B/C, and positrons data \cite{WeinrichEtAl2020,WeinrichEtAl2020b}.

For DM searches in the GeV-TeV energy range, the halo size boundary $L$ and spatial diffusion play the main role for light antinuclei \cite{MaurinEtAl2006}, supplemented by bremsstrahlung~\cite{CirelliEtAl2013}, synchrotron or inverse Compton energy losses for positrons. The role of these parameters is highlighted in the next \citesec{sec:primaries} where we pedagogically derive approximate analytical expressions for the primary fluxes. Complicated aspects regarding CR propagation at low rigidity (combination of convection, reacceleration, low-rigidity break in the diffusion coefficient, solar modulation, etc.) remain important though, as shown in \citesec{sec:stats}.

\subsection{Dark matter source distribution}

For annihilating DM, the appropriate source term ${\cal Q}$ featuring in \citeeq{eq:prop} is given (at some position $\vec{x}_s$) by
\ben
\label{eq:dm_source}
{\cal Q}_{\chi,a}(E,\vec{x}_s) &=& {\cal S}_\chi^\odot \,\frac{dN_a(E)}{dE}\,\left( \frac{\rho(\vec{x}_s)}{\rho_\odot}\right)^2\,,
\een
where $dN_a/dE$ is the spectrum of CRs of species $a$ injected by self-annihilation process, $\rho(\vec{x}_s)$ is the mass density profile, $\rho_\odot$ is the DM density at the position of the solar system, and
\ben
{\cal S}_\chi^\odot \equiv \xi\frac{\sigv}{2} \left(\frac{\rho_\odot}{m_\chi}\right)^2\,.
\een
Above, \mchi\ is the DM particle mass, \sigv\ the thermally averaged annihilation cross section (times speed), $\xi =1$ (or 1/2) if DM particles are (not) self-conjugate, $\rho(\vec{x}_s)$ is the mass density profile of the dark halo.

Such profile is expected to behave approximately like $r^{-\gamma}$ in the inner parts of the Milky Way within the solar circle, where $r$ is the Galactocentric distance and $\gamma\sim[0,1]$ (\ie~between a core and a cusp) \cite{Zhao1996,NavarroEtAl1996a,MerrittEtAl2005}---this is consistent with current kinematic data \cite{McMillan2017,CautunEtAl2020}. Therefore the source may strongly intensify toward the Galactic center, which is located at a distance of $\Rsun\sim 8.2$~kpc from the solar system \cite{GravityCollab2019}, and where a very hot spot of DM annihilation lies in the case of a cuspy halo (see next section).
For definiteness, our illustrations of the \min{}-\med{}-\max{} fluxes are based on an NFW profile \cite{NavarroEtAl1996a}, $\rho(r) = \rho_s (r/r_s)^{-1}(1+r/r_s)^2$, where $\rho_s$ is the scale DM density, but our benchmark transport parameters apply to any profile.

%
\section{Dependence of DM-produced primary CRs on propagation parameters}
\label{sec:primaries}

Diffusion occurs in a magnetic slab of half-height $L$, beyond which magnetic turbulences decay away, leading to CR leakage (free streaming). Therefore, $L$ characterizes a specific spatial scale beyond which CRs can escape from the Milky Way. One can easily deduce that an important consequence for the DM-produced CR flux is related to $L$ itself \cite{TailletEtAl2003,MaurinEtAl2003}, and to
the possible hierarchy between $L$ and \Rsun. In particular, $L\lesssim \Rsun$ leads to a flux approximately set by the local DM density, and $L\gtrsim \Rsun$ leads to an additional important contribution from the Galactic center. From this very simple argument, it is already clear that $L$ will play an important role in defining parameter sets that maximize or minimize the DM-produced CR flux, as we will see below.

Note that since $L$ is usually found to be much smaller than the typical scale radius $r_s\sim 20$~kpc of the DM halo \cite{McMillan2017,CautunEtAl2020}, we approximate (for this section only) the DM density profile by
\ben
\rho(r) \simeq \rho_\odot\left(\frac{r}{R_\odot}\right)^{-\gamma}\,.
\label{eq:rhodm}
\een

We derive below, under simplifying assumptions, the dependence of antiprotons and positrons exotic signals on the halo size $L$ and the diffusion coefficient $K(R)$. We highlight in particular the specific dependence of the Galactic center hotspot source term. Note that the following approximate analytical results are not meant to be compared with accurate predictions, but rather to highlight the decisive role of some of the propagation parameters from concrete physical arguments.

\subsection{Light antinuclei}
We start with the case of light antinuclei in general, but stick to antiprotons without loss of generality---arguments similar to those presented below can be found in several past studies, \eg~\cite{MaurinEtAl2002,DonatoEtAl2004,MaurinEtAl2006,BringmannEtAl2007}. 

Since we are interested in the flux prediction above a few GeV, we can approximately describe the antiproton propagation as being entirely of diffusive nature (neither energy loss nor gain). Forgetting for the moment the spatial boundary conditions associated with our slab model, we can start the discussion in terms of the three-dimensional (3D) Green function, derived with spatial boundaries sent to infinity. In that case, the Green function associated with the steady-state propagation equation is simply given by
\ben
{\cal G}_{\bar p}(E,\vec{x}_\odot\leftarrow\vec{x}_s) = \frac{1}{4\,\pi\,K(E)|\vec{x}_\odot-\vec{x}_s|}\,,
\label{eq:gpbar}
\een
where $K(E)$ is the scalar, rigidity-dependent (equivalently energy-dependent) diffusion coefficient. The Green function is related to the probability for an antiproton injected at point $\vec{x}_s$ to reach a detector located at Earth at point $\vec{x}_\odot$. For relativistic antiprotons (with speed $v\sim c$), the flux at the detector position derives from the Green function through
\ben
\frac{d\phi_{\bar p}(E,\vec{x}_\odot)}{dE} = \frac{v}{4\,\pi}\int_{\rm slab} d^3\vec{x}_s\,{\cal G}_{\bar p}(E,\vec{x}_\odot\leftarrow\vec{x}_s)\,{\cal Q}_{\chi,\bar p}(E,\vec{x}_s)\,.\nn\\
\label{eq:pbar_flux}
\een
Here ${\cal Q}_{\chi,\bar p}(E,\vec{x}_s)$ is the source term for DM-induced antiproton CRs, which has been introduced in \citeeq{eq:dm_source}.

Now, let us try to artificially introduce the vertical boundary condition, \ie~the most stringent one, in the form of an absolute horizon of size $\alpha \, L$, where $\alpha$ is a number of order ${\cal O}(1)$. Crudely assuming that the DM density is quasi-constant within the magnetic slab, one can readily integrate \citeeq{eq:pbar_flux} to get
\ben
\frac{d\phi_{\bar p}(E,\vec{x}_\odot)}{dE} &\approx& \frac{v}{4\,\pi} \frac{\alpha^2 L^2}{K(E)}
\left\langle {\cal Q}_{\chi,\bar p}(E,\vec{x}_s) \right\rangle_{V^\odot_L}\nn\\
\Rightarrow \frac{d\phi_{\bar p}(E,\vec{x}_\odot)}{dE} &\propto & \frac{L^2}{K(E)}\,.
\label{eq:dependence_pbar_flux}
\een
Above, $\left\langle\ldots\right\rangle_{V^\odot_L}$ represents an average over a volume $V^\odot_L\propto (\alpha L)^3$ centered at Earth and delineated by an horizon of size $\alpha\, L$. Equation~(\ref{eq:dependence_pbar_flux}) displays the main dependence of the light nuclei fluxes in terms of the main propagation parameters, $\propto L^2/K.$ It can also be easily shown that this same ratio, which has the dimensions of a time, corresponds to the typical residence time of CRs inside the magnetic halo, $\tau_{\rm res}$. The physical interpretation is therefore simple: the flux scales linearly with the time that CRs spend diffusing before escaping.

 An even simpler scaling relation emerges if one adds the information that $K/L$ is strongly constrained by measurements of secondary-to-primary ratios (\eg~\cite{MaurinEtAl2001}). In particular, if the DM density were to be constant, the flux of primary antinuclei would merely scale linearly with $L$ \cite{MaurinEtAl2006}. In reality, except for the case of an extended-core DM halo, the source term is not constant: it can be assumed to scale like $1/r^2$ for DM self-annihilation and in a cuspy NFW halo. Placing oneself in the observer frame ($\vec{x}_{s/\odot}=\vec{x}_s-\vec{x}_\odot$), and taking ${\cal Q}_{\chi,\bar p}^\odot(E)={\cal S}_\chi^\odot dN_{\bar p}(E)/dE$, one can get
\ben
\frac{d\phi_{\bar p}(E,\vec{x}_\odot)}{dE} &\approx& \frac{v}{4\,\pi}\,{\cal Q}_{\chi,\bar p}^\odot (E)\nn\\
&\times&\int_{V^\odot_L} d^3\vec{x}_{s/\odot}\,\frac{\left(\frac{R_\odot}{|\vec{x}_{s/\odot}+\vec{x}_{\odot}|}\right)^2}{4\,\pi\,K(E)|\vec{x}_{s/\odot}|}\nn\\
&\approx& \frac{v\,R_\odot^2}{4\,\pi\,K(E)}\,{\cal Q}_{\chi,\bar p}^\odot (E)\sum_{l=1}^\infty \frac{(\alpha \,L/R_\odot)^{2l}}{2l(2l-1)}\,.
\label{eq:dependence_pbar_flux_general}
\een
This generalizes \citeeq{eq:dependence_pbar_flux}, and indeed the $l=1$ term gives back the scaling relation previously derived.
This series approximation is valid for $\alpha\,L\ll R_\odot$, but it is enough to grasp the growing impact of $L$ when it increases up to the Galactic center distance \Rsun\ . We emphasize that this very simple calculation does qualitatively capture the exact DM phenomenology very well (see \citesec{sec:results}).

Below a few GeV, this simple picture might be altered as other processes become more efficient than spatial diffusion. This can typically occur in the \quaint\ propagation scheme \cite{GenoliniEtAl2019}, in which diffusive reacceleration could be strong without spoiling energetics considerations \cite{DruryEtAl2017}. Diffusive reacceleration redistributes low-energy particles toward slightly higher energy, and can affect the GeV and sub-GeV predictions significantly \cite{DonatoEtAl2004}. 
This occurs when the typical timescale for reacceleration $\propto K(E)/V_{\rm a}^2$ becomes smaller than other timescales, like the disk-crossing diffusion ($\propto hL/K(E)$, where $h$ is the typical ISM disk half-height), or the convective wind ($\propto h/V_{\rm c}$) timescales. Therefore, reacceleration may play a role at low energy when the Alfv\'en speed $V_{\rm a}$ becomes large, which can further be used to maximize the flux of DM-produced antinuclei in this energy range (see \citesec{sec:stats}).

\subsection{Positrons}
We now turn to the case of positrons (identical considerations apply to electrons). We can basically use the same reasoning as for antiprotons/antinuclei, except that at energies above a few GeV, both spatial diffusion and energy losses are important in characterizing their propagation in the magnetic halo. We can again start from the infinite 3D Green function:
\ben
{\cal G}_{e^+}(E,\vec{x}_\odot\leftarrow E_s,\vec{x}_s) = \frac{1}{b(E)}\frac{e^{-\frac{|\vec{x}_\odot-\vec{x}_s|^2}{2\,\lambda^2}}}{(2\,\pi\,\lambda^2)^{3/2}}\,,
\label{eq:gpos}
\een
where we define the energy-dependent positron propagation scale $\lambda$ as
\ben
\lambda^2 \!&=&\! \lambda^2(E,E_s) = 2\int_E^{E_s}dE'\,\frac{K(E')}{b(E')}\nn\\
\Rightarrow \lambda^2 \!&=&\! \frac{2\,K_\star\,\tau_l}{({\omega}\!-\!\delta\!-\!1)}\varepsilon^{-{\omega}+\delta+1}\left(\! 1 - \left(\frac{\varepsilon_s}{\varepsilon}\right)^{-{\omega}+\delta+1}\!\right).
\label{eq:def_lambda}
\een
In the latter step, we have assumed that the diffusion coefficient and the energy loss rate obey single power laws in energy, with $K(\varepsilon)=K_\star \varepsilon^\delta$, and $b(\varepsilon)=(E_\star/\tau_l)\varepsilon^{\omega} $, defining the dimensionless energy $\varepsilon\equiv E/E_\star$, where $E_\star$ is a reference energy (or the energy of interest). Above a few GeV, energy losses are mostly due to synchrotron and inverse Compton losses, with $\omega=2$ and $\tau_l\approx 10^{16}$~s (for $E_\star=1$~GeV) \cite{DelahayeEtAl2010}. In the inertial regime of spatial diffusion, typical values for $\delta$ are currently found around 0.5 \cite{GenoliniEtAl2019,WeinrichEtAl2020,WeinrichEtAl2020b}. Correctly including the spatial boundary conditions would significantly affect the legibility of the exact solution to the transport equation \cite{BulanovEtAl1974,BaltzEtAl1998,LavalleEtAl2007,BoudaudEtAl2017a}, but, like for antinuclei, we can roughly account for the vertical boundary by limiting the spatial integration to the local volume within $\alpha L$, ${\cal V}_L^\odot$. The positron flux is then given by
\begin{align}
\frac{ d\phi_{e^+}(E,\vec{x}_\odot) }{dE}  & = \frac{v}{4\pi} \nn\\
\!\!\!\times\! \int_E^\infty \!\!\!dE_s\!\int_{{\cal V}_L^\odot} \!\!d^3\vec{x}_s\, 
& {\cal G}_{e^+}(E,\vec{x}_\odot\!\leftarrow\! E_s,\vec{x}_s) \,{\cal Q}_{\chi,e}(E_s,\vec{x}_s),
\end{align}
where an integral over energy has appeared, making the phenomenological discussion slightly more tricky than for antinuclei.

Note that if the dimensionless source energy $\varepsilon_s\gg\varepsilon$, then it no longer affects the propagation scale $\lambda$, whose order-of-magnitude estimate, taking $E_\star=1$~GeV, reads
\ben
\lambda^2\approx (4.5\,{\rm kpc})^2\!\!\left(\!\frac{K_\star}{0.05\,{\rm kpc^2/Myr}} \frac{\tau_l}{10^{16}\,{\rm s}} \right) \!\!\left(\frac{E}{10\,{\rm GeV}}\right)^{\!\!-1/2}\!\!\!\!\!\!\!\!\!\!.
\label{eq:eval_lambda}
\een
CR propagation becomes quickly short-range for positron energies above $\sim 10$~GeV, which can be used to simplify the discussion.

Thus, let us consider two different regimes for the propagation scale $\lambda$. When the positron energy $\varepsilon$ is large, or when $\varepsilon\to \varepsilon_s$, we are in the regime of vanishing propagation scale, $\lambda\to 0$. In that case, the Green function simplifies to
\ben
    {\cal G}_{e^+}(E,\vec{x}_\odot\leftarrow E_s,\vec{x}_s)\overset{\lambda\to 0}{\longrightarrow}
    \frac{\delta^3(\vec{x}_\odot-\vec{x}_s)}{b(E)} \ \theta(E_s-E)\,,\,
\een
such that the DM-produced positron flux is readily given by
\ben
\frac{d\phi_{e^+}(E,\vec{x}_\odot)}{dE} \overset{\lambda\to 0}{\longrightarrow} \frac{v\,{\cal S}_\chi^\odot}{4\,\pi\,b(E)}
\int_E^\infty dE_s\,\frac{dN_{e^+}(E_s)}{dE_s}\,.
\label{eq:local_positron_flux}
\een
This result is important, because it implies that at sufficiently high energy, the positron flux no longer depends on propagation parameters related to spatial diffusion, and is not sensitive to spatial boundary effects either. It only depends on the local energy loss rate, such that the prediction uncertainties are essentially set by those in the Galactic magnetic field and in the interstellar radiation field (ISRF)---for details, see \eg~\cite{DelahayeEtAl2010,PorterEtAl2017}.

 In the opposite regime (small positron energy $\varepsilon$), we can proceed as we did for antinuclei, and if $\lambda\gtrsim L$, we can neglect the Gaussian suppression in \citeeq{eq:gpos} such that the DM-annihilation-induced positron flux scales roughly as
\begin{align}
\frac{d\phi_{e^+}(E,\vec{x}_\odot)}{dE} &\approx
\frac{d\phi_{e^+}(E,\vec{x}_\odot)}{dE}\Big|_{\lambda\to 0} \nn\\
+  \frac{v\,{\cal S}_\chi^\odot R_\odot^3}{b(E)} 
 \sum_{l=1}^\infty & \frac{(\alpha L/R_\odot)^{2l+1}}{(2l+1)(2l-1)}
\int_{E_>}^{\infty} \frac{dE_s}{(2\,\pi\,\lambda^2)^{3/2}} \frac{dN_{e^+}(E_s)}{dE_s}
\nn\\
\overset{l=1}{\propto} \, \frac{(\alpha L)^3}{\lambda^3} & \propto\left(\frac{L^2}{K(\varepsilon_>)\tau_l/\varepsilon_>}\right)^{\!3/2} 
\propto\left(\frac{L^2}{K_\star}\right)^{\!3/2}\!\!\!\!\!.
\label{eq:global_positron_flux}
\end{align}
The first term on the right-hand side is the $\lambda\to 0$ limit derived in \citeeq{eq:local_positron_flux} above, corresponding to the subdominant local yield when $E \leftarrow E_s$ in the energy integral, and the second term adds up farther contributions from source energies beyond a critical value of $E_> \gtrsim E$, for which $\lambda$ is assumed to be of order $L$ (corresponding to the GeV energy range, see \citeeq{eq:eval_lambda}). The series expansion is performed supposing a source term scaling like the (squared) density profile given in \citeeq{eq:rhodm} with $\gamma=1$ (\ie~an NFW halo profile), and is formally valid in the regime $L\ll R_\odot$. As for the case of antinuclei (see \citeeq{eq:dependence_pbar_flux}) this helps understand the growing impact of $L$ as it increases (terms $l>1$ needed), which implies collecting annihilation products closer and closer to the Galactic center. The leading $l=1$ term exhibits a dependence in the same combination of transport parameters, $L^2/K_\star$, as for antinuclei (though to a different power). Recall that $K_\star$, the normalization of the simple power-law approximation of the diffusion coefficient, can be traded for the standard $K_0$ normalization of the complete expression given in \citeapp{app:1D2D}; mind also the specific energy dependence.

Finally, we stress that positrons are much more sensitive to low-energy processes than light antinuclei, simply due to their comparatively smaller inertia. They are particularly responsive to diffusive reacceleration, which may strongly affect their low-energy spectra up to $\sim 5$-10~GeV when efficient enough \cite{DelahayeEtAl2009,BoudaudEtAl2017a}. As already mentioned above, some of the propagation configurations we consider, the \quaint\ model for instance, feature a potentially high level of reacceleration, which could imply a transition from energy-loss dominated (at high energy) to reacceleration dominated transport (at low energy) for positrons. In that case, low-energy yields which may originate from the Galactic center are pushed to slightly higher energies, which leads to a significant ``bumpy'' increase of the DM-induced flux. Like for antinuclei, this occurs whenever the typical timescale for reacceleration $\propto K(E)/V_{\rm a}^2$  becomes smaller than the other relevant timescales, naturally selecting large values of the Alfv\'en speed $V_{\rm a}$.

\subsection{Hierarchy of relevant parameters}

We have just shown that DM-produced primary fluxes of antimatter CRs scale like powers of $L^2/K$, i.e. powers of the diffusion time across the magnetic halo; see Eqs.~(\ref{eq:dependence_pbar_flux}), (\ref{eq:dependence_pbar_flux_general}), and (\ref{eq:global_positron_flux}). The parameter $L$ itself plays a special role by enlarging the CR horizon toward the Galactic center hot spot. An exception arises for high-energy positrons (typically $E\gtrsim 10$~GeV), for which only inverse Compton and synchrotron energy losses matter, so that the resulting flux becomes independent of the propagation parameters; see \citeeq{eq:local_positron_flux}. More subtle effects may add up at GeV energies, like diffusive reacceleration. In that case, propagation uncertainties are set by uncertainties in the modeling of the magnetic field and of the interstellar radiation field that enter the energy-loss rate (the larger the loss rate, the smaller the flux). For more details on these uncertainties, see \eg~\cite{DelahayeEtAl2010,PorterEtAl2017}.

Since $L/K$ is strongly constrained ($L/K_0\sim$ constant), the main hierarchy in the exotic flux calculation is therefore set by a selection over $L$. It is necessary to further select on the remaining transport parameters to ensure the consistency of the hierarchy down to the lowest energies. For instance, the \slim\ and \big\ schemes, close to pure diffusion models, have a low-rigidity break index $\delta_{\rm l}$. This parameter is correlated to other transport parameters, and it strongly impacts on the residence time of CRs in the halo, and thereby affects the hierarchy in the flux predictions. Besides, the \quaint\ scheme can allow for a significant amount of reacceleration, which also impacts on the primary fluxes (by pushing upward the low-energy yield and padding the flux at some critical energy that minimizes the relative reacceleration timescale). For these reasons, the dependence of both $L$ and $\delta_{\rm l}$ for \slim\ and \big\, and $L$ and $V_{\rm a}$ for \quaint\ are also used in the following to define our reference models \min, \med, and \max.

\section{Statistical method}
\label{sec:stats}

In this section, we describe the statistical method used to derive the parameters corresponding to the \min{}, \med{}, and \max{} benchmarks. For simplicity, we focus on the \slim{} scheme, which has 5 free parameters, namely $L$, $\delta$, $K_0$, $R_\mathrm{l}$, and $\delta_\mathrm{l}$ (see details in App.~\ref{sapp:models}). The same approach can also be applied to the \big\ and \quaint\ schemes (see App.~\ref{app:min_med_max_general}).

\begin{figure}[t!]
\includegraphics[width=0.9\columnwidth]{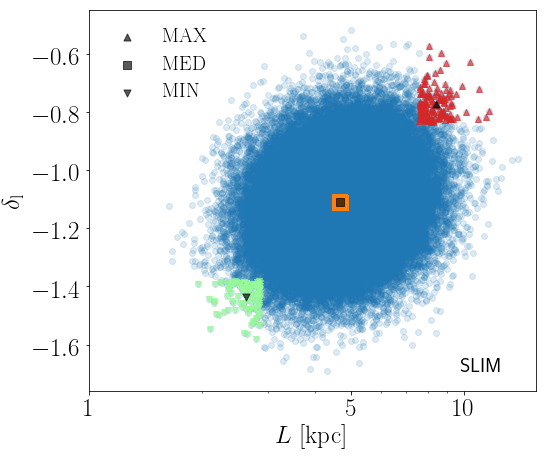}
\caption{
The constellation of blue dots features a sample of $10^{5}$ randomly drawn {\slim} models.
Along each of the directions $\log_{10}{L}$ and $\delta_{\rm l}$, the red and green models are located at more than 2 standard deviations from the mean. The barycenters of these populations, defined with respect to all CR parameters, respectively yield the {\max} and {\min} configurations, depicted by the upward and downward  black triangles.
The {\med} model corresponds to the barycenter of the orange sub-sample. The latter is defined by requiring that the quantiles with respect to $\log_{10}{L}$ and $\delta_{\rm l}$ are equal to the average value $q_{\textrm{\med{}}} = 0.5$ up to a width $p$.
}
\label{fig:Def_MinMedMax_slim}
\end{figure}

Our starting point is the transport parameters derived in \cite{WeinrichEtAl2020b}, from the combined analysis of \ams{} Li/C, B/C, and Be/B data, and existing $^{10}$Be/Be+$^{10}$Be/$^9$Be data (see Table~2 of \cite{WeinrichEtAl2020b}). More specifically, we use the best-fit values and covariance matrix of uncertainties\footnote{The best-fit values and the covariance matrix are given explicitly in App.~\ref{sapp:covmat}. Subtleties and further checks regarding the derivation of these quantities are given in App.~\ref{app:1D2D} and~\ref{app:mcmc}; the latter shows that the parameters follows at first order a multidimensional Gaussian.} to draw a collection of $10^{5}$ {\slim} (correlated) propagation parameters. This sample is displayed in the $(L ,  \delta_{\rm l})$ plot of Fig.~\ref{fig:Def_MinMedMax_slim}. Each blue point stands for a particular model within the \slim{} propagation scheme. The constellation of dots is nearly circular, indicating that the CR parameters $\log_{10}{L}$ and $\delta_{\rm l}$ are not correlated with each other.

To define the \max{} (resp. \min{}) configuration, we start selecting a sub-sample of {\slim} models whose quantiles relative to $\log_{10}{L}$ and to $\delta_{\rm l}$ are both larger (resp. smaller) than a critical value of
\ben
q_{^{\rm \max}_{\rm \min}} = \frac{1}{2} \left( 1 \pm \mathrm{erf} \left(\frac{n}{\sqrt{2}} \right) \right) ;
\een
where $\mathrm{erf}(x) = 2/\sqrt{\pi}\int_{0}^x e^{-t^2}\mathrm{d} t$ is the error function. Along each of the directions $\log_{10}{L}$ and $\delta_{\rm l}$, these models are located at more than $n$ standard deviations from the average configuration. We show in the next section that a value of $n=2$ efficiently brackets the DM-produced primary fluxes, whatever the annihilation channel. In Fig.~\ref{fig:Def_MinMedMax_slim}, this sub-sample corresponds to the red dots lying in the upper-right corner (resp. green points in the lower-left corner) of the blue constellation. Once this population has been drawn, the {\max} (resp. \min{}) model is defined as its barycenter inside the multi-dimensional space of all CR parameters. It is identified by the upward (resp. downward) black triangle.

For the {\med} model, we proceed slightly differently. The orange square in Fig.~\ref{fig:Def_MinMedMax_slim} corresponds to a sub-sample of models whose quantiles relative to $\log_{10}{L}$ and to $\delta_{\rm l}$ are both in the range extending from $q_{\textrm{\med{}}} - {p}/{2}$ to $q_{\textrm{\med{}}} + {p}/{2}$, with $q_{\textrm{\med{}}} = 0.5$ and $p$ a `width parameter' specified below. Once that population is selected, the {\med} model corresponds once again to its barycenter configuration in the multi-dimensional space of all CR parameters. It is shown as a black square lying at the center of the orange square. The parameters of the {\med} model have been derived with $p=0.03$. However, Fig.~\ref{fig:Def_MinMedMax_slim} has been made using $p=0.1$ for legibility, as a smaller value would have shrinked the orange zone underneath the triangle.

\begin{figure}[t!]
\includegraphics[width=0.7\columnwidth]{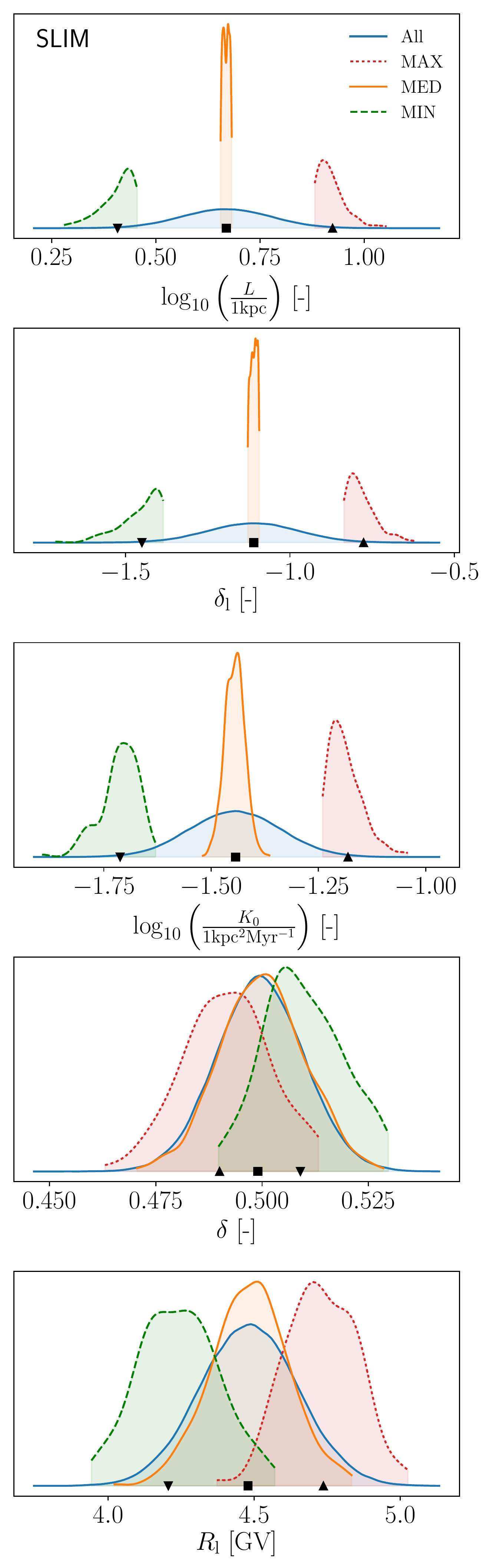}
\caption{
Probability distribution functions for the {\slim} propagation parameters. The blue curves correspond to the full collection ({\em All}) of $10^{5}$ randomly drawn {\slim} models. The green-dotted, orange-dashed, and red-solid distributions stand respectively for the {\min}, {\med}, and {\max} sub-samples, and the symbols correspond to the determined benchmark values (see text).
}
\label{fig:stat_pars}
\end{figure}
The probability distribution functions (PDFs) of the {\slim} propagation parameters have been extracted for the entire population of $10^{5}$ randomly drawn models. They are represented in Fig.~\ref{fig:stat_pars} by the blue curves labeled {\em All}. Similar PDFs have also been derived for the {\min}, {\med} and {\max} sub-samples. They respectively correspond to the green-dotted, orange-dashed, and red-solid lines. As in Fig.~\ref{fig:Def_MinMedMax_slim}, we have used a width of $p=0.1$ to show the PDFs for the {\med} population.
We first notice that the blue PDFs ({\em All}) extend broadly over the entire accessible range of propagation parameters. They correspond to the global sample.
This is not quite the case for the PDFs relative to the {\min}, {\med}, and {\max} sub-samples. These populations have been extracted by selecting particular values of $\log_{10}{L}$ and $\delta_{\rm l}$. It is therefore no surprise if the corresponding PDFs are quite narrow and well separated from each other in the two upper panels of Fig.~\ref{fig:stat_pars}.
The PDFs relative to the normalization $K_{0}$ of the diffusion coefficient are shown in the middle panel. As said previously, a correlation between $\log_{10}K_0$ and $\log_{10}L$ directly arises from calibrating propagation on secondary-to-primary ratios --- this appears explicitly in \citefig{fig:app_MCMC} in \citeapp{app:mcmc}. This translates into fairly peaked PDFs for $K_{0}$. The {\min} and {\max} models are thus respectively characterized by lower and larger values for $L$, $\delta_{\rm l}$ and $K_{0}$.
The separation induced by a selection of \min-\med-\max\ models from $L$ and $\delta_{\rm l}$ is less striking in the PDFs of the inertial diffusion index $\delta$ and of the position of the low-rigidity break $R_{\rm l}$, but is still observed (they actually slightly correlate with $K_0$, and more strongly with $\delta_{\rm l}$). Our selection procedure allows us to fully account for these slighter correlations, even if these latter parameters have much less impact on the primary fluxes.
In each panel, we finally notice that the blue and orange PDFs have the same mean. The {\med} configuration is actually defined as the barycenter of a sub-population of models selected for their average values of $\log_{10}{L}$ and $\delta_{\rm l}$. This sub-sample sits in the middle of the entire population.

\begin{table}[t]
\caption{
Propagation parameters for the {\min}, {\med}, and {\max} benchmarks for {\slim}.}
\label{tab:stat_parvalues}
\begin{tabular}{p{1cm}p{1.1cm}p{1.1cm}p{2.1cm}p{1.2cm}p{1.2cm}}
\hline\hline
{\slim} & $L$  	    & $\delta$   & $\log_{10} K_0$			& $R_\mathrm{l}$  	& $\delta_\mathrm{l}$ \\
 	     & [kpc]       & 		     &  [kpc$^2$\,Myr$^{-1}$] &  [GV] 		& 		 \\
\hline
{\max} &     8.40     &     0.490     &     -1.18     &     4.74     &     -0.776 \\
{\med} &     4.67     &     0.499     &     -1.44     &     4.48     &     -1.11 \\
{\min} &     2.56     &     0.509     &     -1.71     &     4.21     &     -1.45 \\
\hline
\end{tabular}
\end{table}
The sets of values of the propagation parameters for the {\min}, {\med} and {\max} benchmarks are listed in Table~\ref{tab:stat_parvalues}. Similar tables for the {\big} and {\quaint} cases are shown in App.~\ref{app:min_med_max_general}.

\section{New Min-Med-Max fluxes on selected examples}
\label{sec:results}

Equipped with the samples derived in Sec.~\ref{sec:stats}, we compute numerically some primary fluxes to check if the half-height $L$ of the magnetic halo efficiently gauges them, as proposed in Sec.~\ref{sec:primaries}.
We first derive the positron flux for some representative annihilation channels. To do so, we use the so-called pinching method in a 2D setup~\cite{BoudaudEtAl2017a}, which is the most up-to-date semi-analytical procedure to incorporate all CR transport processes. The DM halo profile is borrowed from~\cite{McMillan2017} with a galacto-centric distance $\Rsun$ of $8.21$~kpc, a local DM density $\rho_\odot$ of $0.383 \; {\rm GeV \, cm^{-3}}$ and a scale radius $r_s$ set to $18.6$~kpc. To ensure a fast convergence of the Bessel series expansion, the central divergence is smoothed according to the method detailed in Sec.~{II.D} of \cite{DelahayeEtAl2008} with a renormalization radius of 0.1~kpc. For definiteness, we use the thermal cross-section $\langle\sigma v \rangle_{\rm th}=3 \times 10^{-26} \; {\rm cm^{3} \, s^{-1}}$, but this parameter can be factored out and is not relevant to our analysis.

\begin{figure}[t]
{\includegraphics[width=0.9\columnwidth]{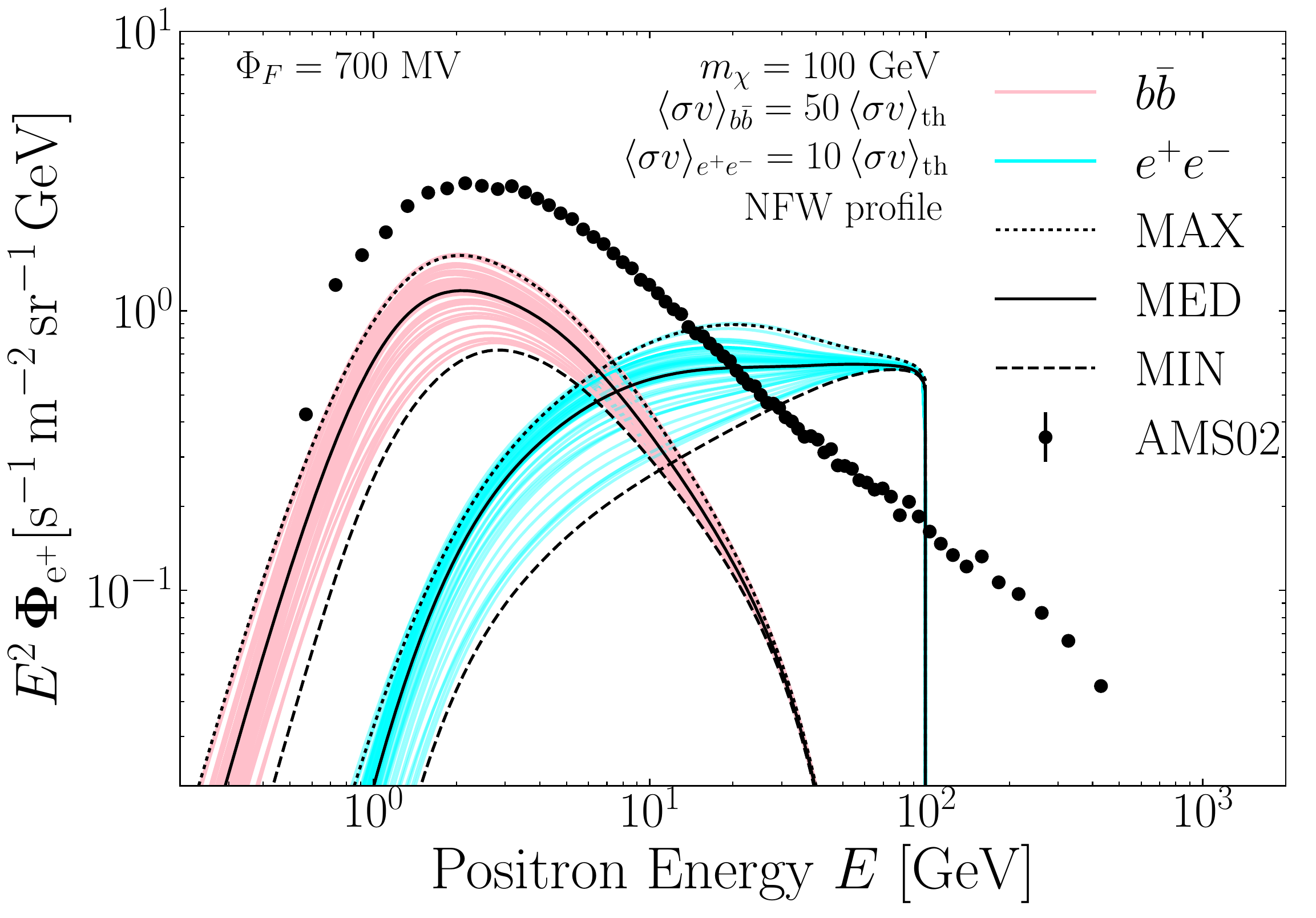}
\includegraphics[width=0.9\columnwidth]{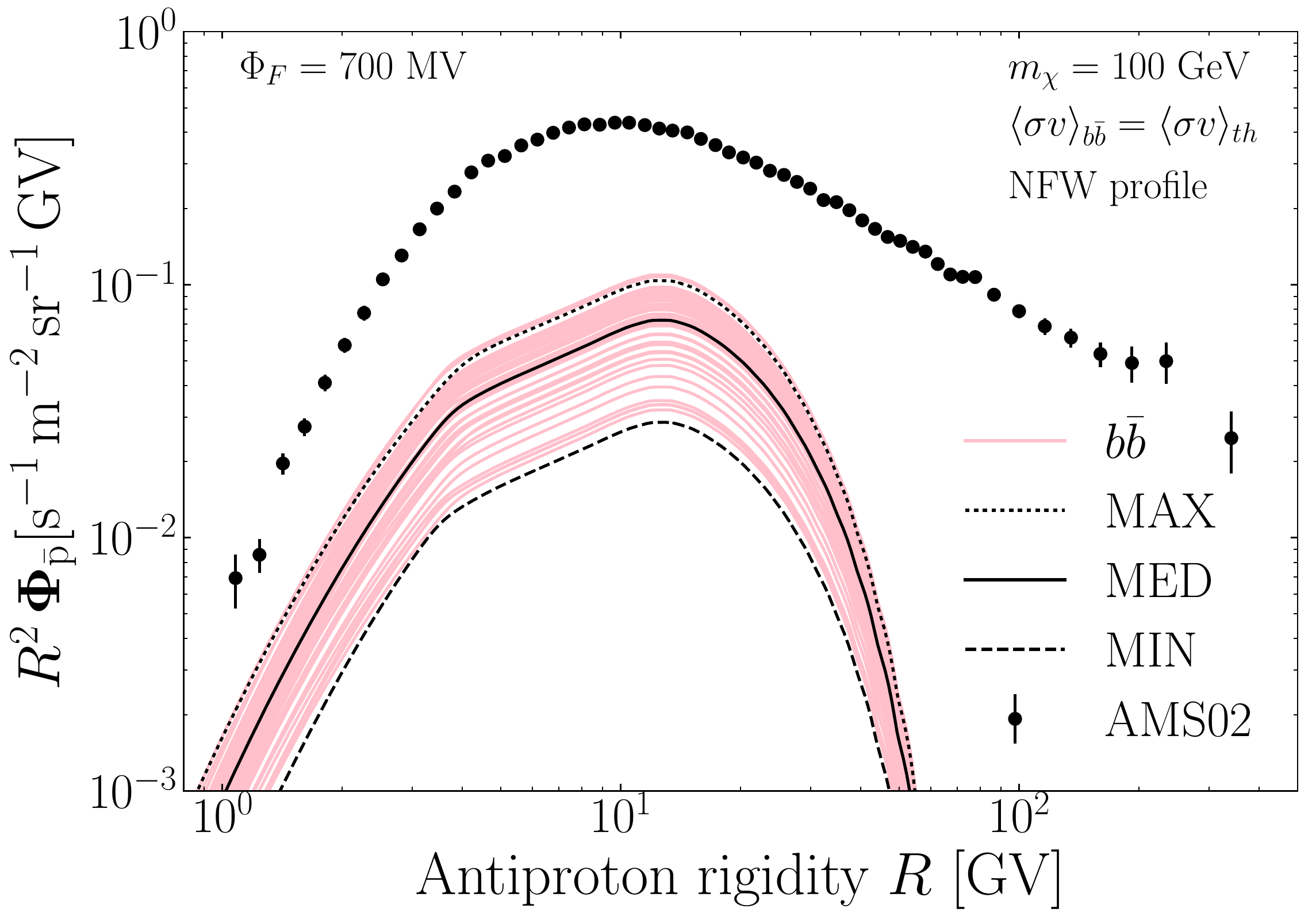}}
\caption{
In the {\bf upper panel}, the primary positron fluxes are plotted as a function of positron energy for two different annihilation channels, i.e. $b \bar{b}$ in pink and $e^{+}e^{-}$ in blue. The annihilation cross section has been respectively set to $1.5 \times 10^{-24}$ and $3 \times 10^{-25} \; {\rm cm^{3} \, s^{-1}}$, to obtain primary fluxes roughly at the same level as the {\sc Ams-02} data~\cite{AguilarEtAl2019}, just for illustration purposes.
The {\bf lower panel} features the antiproton yield for the same DM species and $b \bar{b}$ channel as above, with thermal annihilation cross section.
For each channel, 50 CR models have been randomly selected and drawn in color. The {\min}, {\med}, and {\max} configurations respectively correspond to the dashed, solid and dotted black curves. All fluxes are modulated, with a Fisk potential of $\Phi_F=700\,$MV.
}
\label{fig:flux_ee_bb_fast}
\end{figure}

\begin{figure*}[t]
\includegraphics[width=0.9\textwidth]{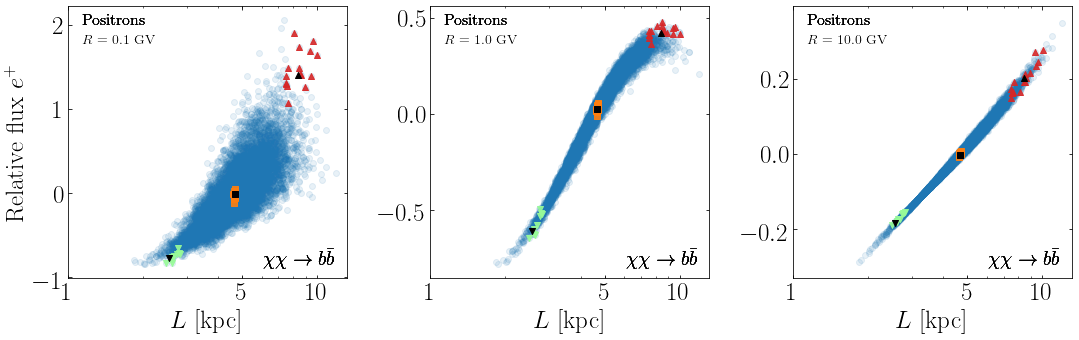}
\includegraphics[width=0.9\textwidth]{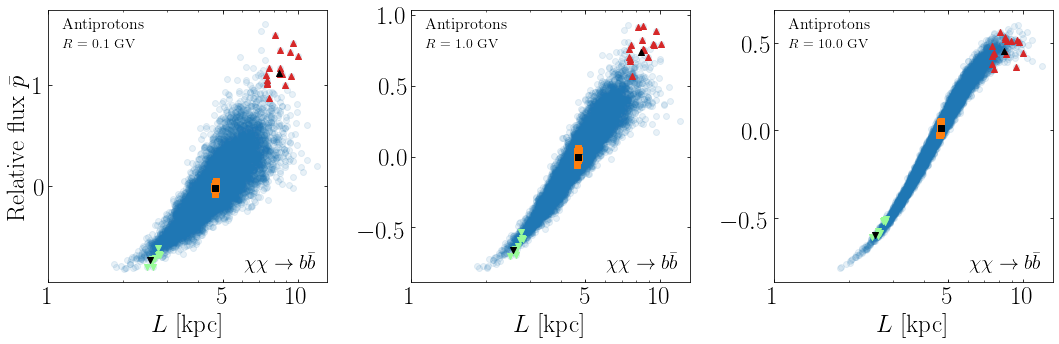}
\caption{
The relative spread of primary fluxes is plotted as a function of the half-height $L$ of the magnetic halo, at three different rigidities. Here we considered a DM particle of mass 100~GeV that annihilates through the $b \bar{b}$ channel. Positrons and antiprotons respectively correspond to the upper and lower panels. Solar modulation is included with a Fisk potential of $\Phi_F=700\,$MV.
In each panel, the green, orange and red points respectively correspond to the {\min}, {\med}, and {\max} samples selected, as explained in section~\ref{sec:stats}. For illustrative purpose, in this figure, $10^{4}$ {\slim} models are randomly drawn (as opposed to the larger ensembles we use in the main analysis), so each colored sample contains around a dozen models. The black squares and triangles indicate the actual loci of the {\med}, {\min} and {\max} configurations.
}
\label{fig:phi_pos_rel_vs_L_bb_at_3_rigidities}
\end{figure*}

\subsection{Fluxes}
In the upper panel of \citefig{fig:flux_ee_bb_fast}, we present a first example of such calculations. The primary positron flux is calculated for a subset of the {\slim} models derived in \citesec{sec:stats} and a DM mass of 100~GeV. The pink and blue curves respectively correspond to $b \bar{b}$ and $e^{+}e^{-}$ channels. The injection spectra ${dN_{e^{+}}}/{dE_s}$ are taken from an improved version of PPPC4DMID~\cite{CirelliEtAl202x}.\footnote{With respect to the original 2010 one \cite{CirelliEtAl2011}, this version is based on an updated release of the collider Monte Carlo code {\sc Pythia} \cite{SjostrandAndSkands2004}, which includes in particular up-to-date information from the LHC runs and an almost complete treatment of electroweak radiations. For all the practical purposes of the examples presented in the section, however, the differences are negligible.}

For positrons, we first notice that whatever the CR model, the predictions for a given annihilation channel converge to the same value at high energy. When the positron energy $E$ is close to the DM mass, the propagation scale $\lambda$ is much smaller than both $L$ and $\Rsun$. As predicted in \citesec{sec:primaries}, the positron flux is then given by \citeeq{eq:local_positron_flux} and is not impacted by diffusion or the magnetic halo boundaries.
Moving toward smaller positron energies, the various curves separate from each other while keeping their respective positions down to approximately 1~GeV. At even lower energies, they are intertwined with one another and the high-energy ordering of the primary fluxes is lost. In the case of the {\slim} parametrization, the low-energy index $\delta_{\rm l}$ of the diffusion coefficient comes into play and redistributes the fluxes in the sub-GeV range.
However, because the {\min}, {\med}, and {\max} models (represented by the black lines) have actually been selected from both $L$ and $\delta_{\rm l}$, they do {\em not} exhibit that trend and the corresponding fluxes follow the expected hierarchy. In particular, the extreme {\min} (dashed) and {\max} (dotted) curves nicely encapsulate the bulk of flux predictions down to the lowest energies. Although they have been derived from CR parameters alone, the {\min} and {\max} configurations can thus be used to determine the range over which primary positron fluxes are expected to lie. This was actually expected from the discussion of \citesec{sec:primaries}.

In the lower panel of \citefig{fig:flux_ee_bb_fast}, the antiproton flux (calculated with the \usine{} code) is derived for the same $b \bar{b}$ channel as for the positrons, but here with a thermal annihilation cross section. This time, diffusion alone dominates over the other CR transport processes. Consequently, whatever the energy, the antiproton flux scales like ${L^{2}}/{K}$, which boils down to $L$ insofar as the ratio ${L}/{K}$ is fixed by B/C data. We notice that the pink curves, which can be considered as a representative sample of all possible antiproton flux predictions, are once again contained within the band delineated by the {\min} (dashed) and {\max} (dotted) lines. The width of this band is furthermore independent of energy and corresponds to a factor of $\sim$ 3.

\subsection{Relative variations and correlations}
In the upper panels of Fig.~\ref{fig:phi_pos_rel_vs_L_bb_at_3_rigidities}, the relative variations of the positron flux $\{ \phi_{e^{+}} - \langle \phi_{e^{+}} \rangle \}/{\langle \phi_{e^{+}} \rangle}$ are plotted as a function of the half-height $L$ for a population of $10^{4}$ {\slim} configurations drawn as in Sec.~\ref{sec:stats}. Each panel corresponds to a different positron rigidity. The flux ${\langle \phi_{e^{+}} \rangle}$ is the population average at that rigidity. Each blue dot represents a different CR model. Fluxes are derived for the $b \bar{b}$ channel. We first notice a clear correlation between the positron flux and the half-height $L$ at 1 and 10~GV. The points are aligned along a thin line and feature the expected increase of $\phi_{e^{+}}$ with $L$.
In the upper-left panel, the same trend is visible but the distribution of blue dots significantly broadens for large values of $L$. This can be understood as follows. At low rigidity, positrons lose energy mostly in the Galactic disk and propagate like nuclei, albeit with a much larger energy loss rate. If the magnetic halo is thin, the CR horizon shrinks with $L$ and the positron flux at Earth has a local origin. At 0.1~GV for instance, energy losses dominate over other CR processes and $\phi_{e^{+}}$ is well approximated by \citeeq{eq:local_positron_flux}; its variance is small.
Conversely, if $L$ is large, the positron horizon reaches the Galactic center which substantially contributes now to the flux. This non-local contribution sensitively depends on the ratio $L^2/K_{\star}$, where $K_{\star}$ stands for the diffusion coefficient at the energy at which $\phi_{e^{+}}$ is calculated. Here, $K_{\star}$ is taken at 0.1~GV and strongly depends on the low-energy parameter $\delta_{\rm l}$. At sub-GeV energies, the ratio ${L}/{K_{\star}}$ is less constrained by the B/C ratio than in the GeV range, hence a large variance which translates into the observed broadening of the flux predictions.
 
The same behavior is observed in the lower panels of Fig.~\ref{fig:phi_pos_rel_vs_L_bb_at_3_rigidities} devoted to antiprotons. At moderate and high rigidities, the relative variations of the antiproton flux $\{ \phi_{\bar{p}} - \langle \phi_{\bar{p}} \rangle \}/{\langle \phi_{\bar{p}} \rangle}$ are nicely correlated with the half-height $L$ as expected from the discussion of Sec.~\ref{sec:primaries}. At low rigidity, the same reasoning as for positrons can be applied to antiprotons. In the sub-GeV region, energy losses mildly dominate over diffusion. Antiprotons are mostly produced locally and their flux at Earth depends on their energy loss rate, especially if the magnetic halo is thin. However, for large values of $L$, the Galactic center with its dense DM distribution becomes visible. Diffusion starts to compete with energy losses. The flux increases like $L^2/K_{\star}$, where $K_{\star}$ is dominated by the low energy parameter $\delta_{\rm l}$. Like for positrons, the variance of the antiproton flux predictions increases, hence the observed broadening of the blue population when $L$ is large.
 
\begin{figure}[t]
\includegraphics[width=0.7\columnwidth]{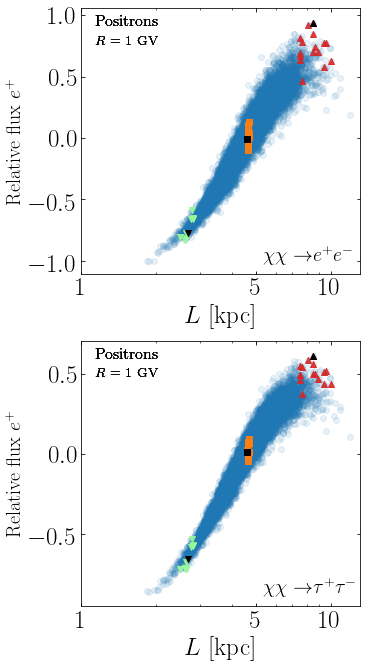}
\caption{
We observe the same trend as in the previous figure for positrons produced from a DM species annihilating into $e^{+}e^{-}$ (left) and $\tau^{+}\tau^{-}$ (right) pairs. The positron fluxes are taken at 1~GV.
}
\label{fig:phi_pos_rel_vs_L_ee_tautau_at_1_GV}
\end{figure}

In each panel of Fig.~\ref{fig:phi_pos_rel_vs_L_bb_at_3_rigidities}, the green, orange and red dots respectively correspond to the {\min}, {\med}, and {\max} sub-samples selected as explained in section~\ref{sec:stats}, i.e. taking $n=2$ and $p=0.03$ for both parameters $L$ and $\delta_{\rm l}$. The green and red populations lie at the lower-left and upper-right boundaries of the constellation of blue dots, while the orange subset sits in the middle.
In Fig.~\ref{fig:phi_pos_rel_vs_L_ee_tautau_at_1_GV}, the same behavior is observed for positrons produced by a DM species annihilating through the $e^{+}e^{-}$ (left) and $\tau^{+}\tau^{-}$ (right) channels.

\begin{figure*}[t]
{\includegraphics[width=0.66\textwidth]{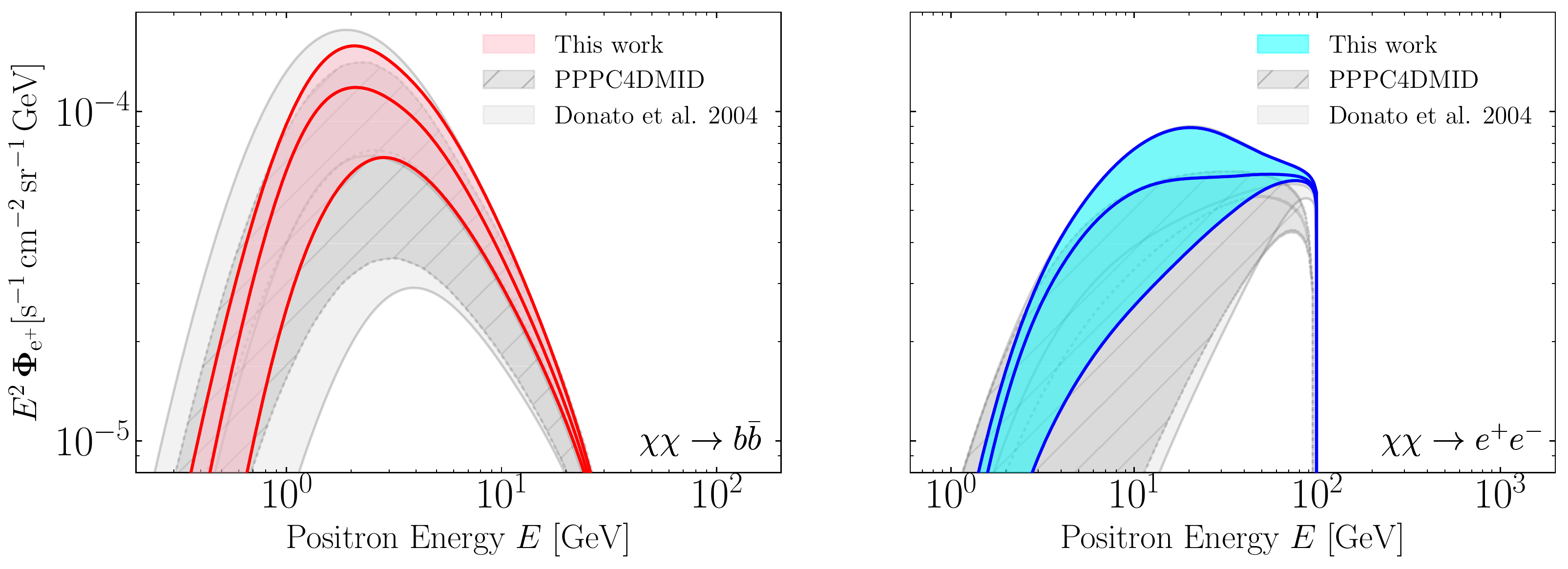}
\includegraphics[width=0.33\textwidth]{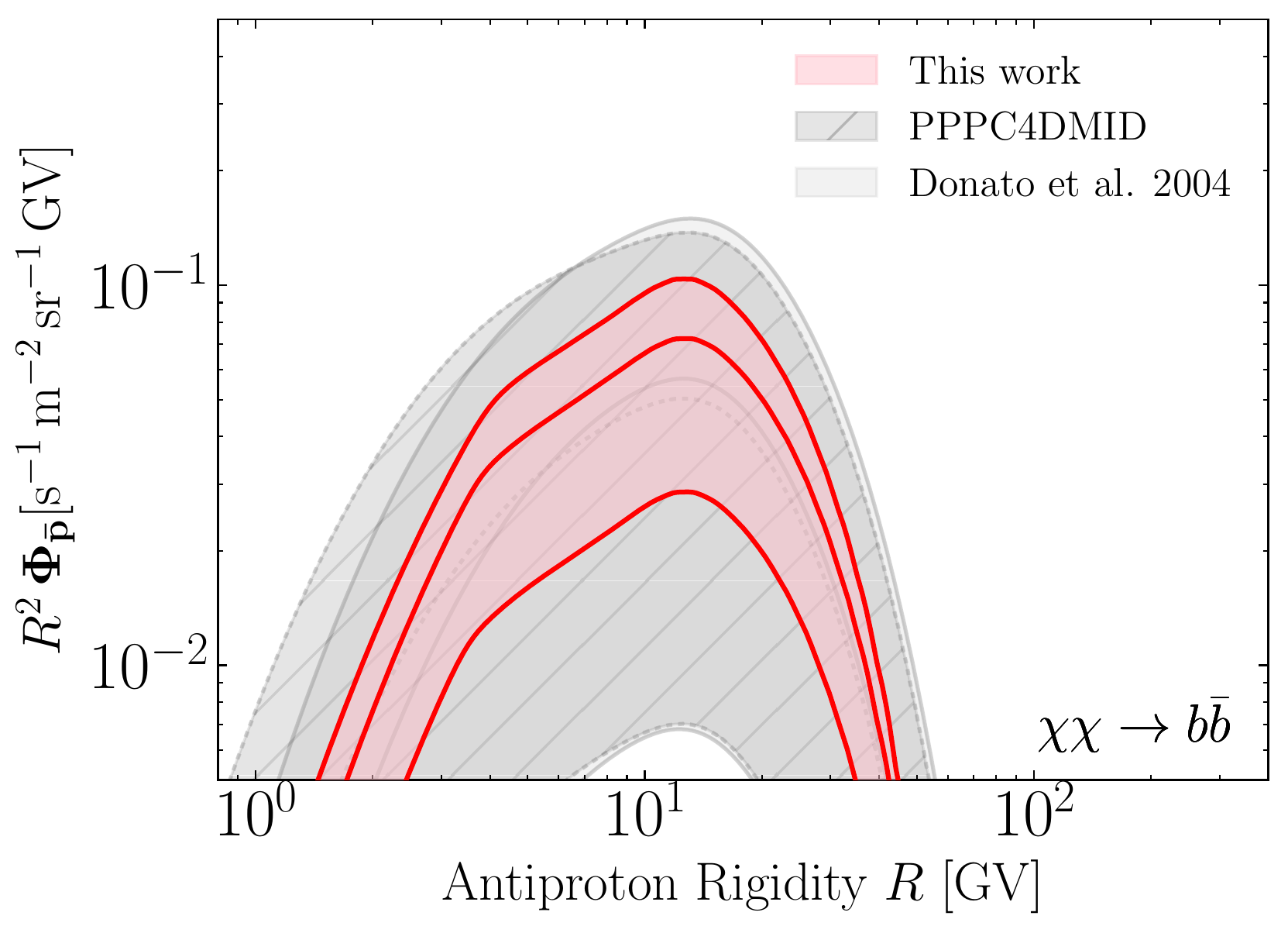}}
\caption{
The theoretical uncertainty on primary fluxes owing to CR propagation has been shrinking as a result of more accurate measurements. The light-gray bands correspond to the original determination of the {\min}, {\med}, and {\max} models by \cite{DonatoEtAl2004} for antiprotons and by~\cite{DelahayeEtAl2008} for positrons. The hatched-Gray regions feature the slightly improved predictions proposed in the framework of PPPC4DMID by~\cite{BoudaudEtAl2014} for antiprotons and~\cite{Buch:2015iya} for positrons. The results of this work are illustrated in the {\slim} case by the pink ($b \bar{b}$ channel) and blue ($e^{+}e^{-}$) strips, for positrons (left and middle panel) and for antiprotons (right panel). They point toward a dramatic improvement of how DM induced fluxes are now calculated.
}
\label{fig:comparisons_min_med_max_old_vs_new}
\end{figure*}

\subsection{Comparison with previous \min{}-\med{}-\max{}}

The {\min}, {\med}, and {\max} models allow to gauge the uncertainty arising from CR propagation. As the precision of CR measurements has been considerably improving in the past decade, so has the accuracy of the theoretical predictions. This trend is clear in Fig.~\ref{fig:comparisons_min_med_max_old_vs_new} where several uncertainty bands are featured for the $b \bar{b}$ and $e^{+}e^{-}$ channels. The light-gray bands correspond to the original analysis by~\cite{DonatoEtAl2004} for antiprotons and by~\cite{DelahayeEtAl2008} for positrons. As mentioned in Sec.~\ref{sec:intro}, the corresponding  configurations were derived by inspecting the behavior of primary antiprotons. A slightly more refined version of the {\min}-to-{\max} uncertainties was proposed in the framework of PPPC4DMID by~\cite{BoudaudEtAl2014} for antiprotons and by~\cite{Buch:2015iya} for positrons. They are represented as hatched-gray regions. The latest determinations, derived in the present work, lie within the pink ($b \bar{b}$) and blue ($e^{+}e^{-}$) strips, for positrons (left and middle panel) and for antiprotons (right panel). These are significantly less extended than in the past, hence a dramatic improvement of how DM induced fluxes are currently determined.
Note that an additional difference for positrons comes from the fact that different prescriptions for energy losses were used in these works. This is seen at high energy where all curves would converge, should energy losses be the same.

\section{Summary and conclusion}
\label{sec:concl}

The propagation in the Galactic environment of the CRs produced by DM annihilations or decays in the diffusion halo is a source of significant uncertainty. Reliably and consistently estimating this uncertainty represents a crucial step toward a better understanding of the potential of indirect detection to constrain the DM properties, and ultimately to the possible discovery of a DM signature in CRs. This is particularly important in light of the recent harvest of accurate CR data, which are opening the way to precision DM searches. These same data allow to pin down, with unprecedented accuracy, the different aspects of CR propagation.

In this spirit, in this paper we have derived new \min{}, \med{}, and \max{} benchmark parameter sets that correspond to the minimal, median, and maximal fluxes of DM-produced CRs in the Milky Way (as allowed by constraints set by standard CRs). They replace their former version, previously used in the literature for antiprotons and positrons. The new derived parameters are actually valid for both species, and for light anti-nuclei more generally, scaling down the uncertainty by a factor of $\sim 2$.
We have worked in the framework of the state-of-the-art Galactic propagation schemes \slim, \big, and \quaint. 
The \min-\med-\max{} parameters for \slim{} are given in \citetab{tab:stat_parvalues} and represent the main output of our work. For convenience we also provide parametric fits for the associated secondary astrophysical predictions in App.~\ref{sapp:pbar_formula}. The corresponding sets for \big{} and \quaint{} (see App.~\ref{app:min_med_max_general}) are given in the form of ancillary files. In practice, the DM practitioner interested in estimating, in an economical and effective way, the variability of DM CR fluxes induced by Galactic propagation can use the \slim{} \min-\med-\max{} sets. For a more complete analysis, the user can also use the \big{} version (the \big{} scheme retains the full complexity of the transport process, with little approximations) and the \quaint{} one (the \quaint{} scheme puts the accent on reacceleration and convection). Going beyond the \min-\med-\max{} references can also be achieve in more involved analyses using the covariance matrices of the propagation parameters provided in App.~\ref{sapp:covmat}.

The computation of CR propagation itself can be performed with semi-analytic codes such as \usine, or, of course, via a dedicated numerical or semi-analytical CR propagation work. In the future, these new benchmarks  configurations will also be implemented in ready-to-use, DM-oriented numerical tools such as the PPPC4DMID.

Our revised \min-\med-\max\ parameter sets lead to significant changes. As illustrated in the several examples above, they reduce by a factor $\cal{O}$(10) the width of the uncertainty band. Hence any new DM ID analysis employing these new sets can be expected to reduce the uncertainty of the DM properties (most notably the constraints on the annihilation cross section or the decay rate) by the same factor.

\begin{acknowledgments}
  This work has been supported by Universit\'e de Savoie, appel \`a projets: {\em Diffusion from Galactic High-Energy Sources to the Earth (DIGHESE)}, by the national CNRS/INSU PNHE and PNCG programs, co-funded by INP, IN2P3, CEA and CNES, and by Villum Fonden under project no.~18994. We also acknowledge financial support from the ANR project ANR-18-CE31-0006, the OCEVU Labex (ANR-11-LABX-0060), from the European Union's Horizon 2020 research and innovation program under the Marie Sk\l{}odowska-Curie grant agreements N$^\circ$ 690575 and N$^\circ$ 674896 and from the {\sc Cnrs} $80|${\sc Prime} grant scheme (`{\sc DaMeFer}' project). M.C.~acknowledge the hospitality of the Institut d'Astrophysique de Paris ({\sc Iap}), where part of this work was done.
\end{acknowledgments}

\appendix

\section{Propagation model, parameters, and ancillary files}
\label{app:1D2D}

This Appendix is devoted to a more thorough description of the propagation formalism used in the main text, and its individual elements.

The ingredients of the 2D propagation model are recalled in Sect.~\ref{sapp:models}. The propagation parameter values are taken from the 1D model analysis of \cite{WeinrichEtAl2020,WeinrichEtAl2020b}, and we show in Sect.~\ref{sapp:1Dto2D} why we can adopt them for our 2D model too. For readers who wish to go beyond the \min-\med-\max{} parameter sets for their analyses, we provide in Sect.~\ref{sapp:covmat} the best-fit and covariance matrix of uncertainties on the parameters for various configurations, and specify how to draw from them in Sec.~\ref{sapp:draw_params}. We also provide in Sect.~\ref{sapp:pbar_formula} a parameterized formula for the secondary $\bar{p}$ and $e^+$ fluxes, that should prove useful for readers who wish to study the DM contribution together with a reference background calculation.

\subsection{2D thin disk/thick halo model, diffusion coefficient, and  configurations}
\label{sapp:models}

The general steady-state transport equation has been introduced above in \citeeq{eq:prop}. Following~\cite{MaurinEtAl2001,DonatoEtAl2004}, CRs propagate and are confined in a cylindrical geometry of half-thickness $L$ (diffusive halo) and radius $R$. Standard CR sources and the gas are pinched in an infinitely thin disk of half-thickness $h=100~{\rm pc} \ll L$), whereas exotic sources are distributed following, for instance, the DM distribution. Only sources inside the diffusive halo are considered here, since it was shown (see App.~B of \cite{BarrauEtAl2002}) that sources outside have a negligible contribution---see however \cite{PerelsteinEtAl2011} for calculations with a position-dependent diffusion coefficient leading to estimates of $\sim 25\%$ of the total.

For transport, the model assumes (i) a constant convection term $V_{\rm c}$ perpendicular to the disk (positive above and negative below), (ii) an isotropic and homogeneous diffusion coefficient with a broken power-low at low (index $l$) and high (index $h$) rigidity,
\ben
  \label{eq:def_K}
  \!\!\!\!K(R) \!=\! {\beta^\eta} K_{0} \!\!
  \left[ 1 \!+\!\! \left( \! \frac{R_{\rm l}}{R} \right)^{\!\!\!\frac{-\delta_{\rm l}+\delta}{s_{\rm l}}} \right]^{\!\!s_{\rm l}}
  \!\!\!\!\!{\left( \! \frac{R}{1\,{\rm GV}\!\!} \right)^{\!\!\delta}}
  \!\!\left[  \!1 \!+ \!\! \left( \! \frac{R}{R_{\rm h}} \right)^{\!\!\!\frac{\Delta_{\rm h}}{s_{\rm h}}}
    \!\!\right]^{\!\!-s_{\rm h}}\!\!\!\!\!\!\!\!\!,\!\!\!\!
\een
and (iii) a diffusion in momentum space \cite{SeoEtAl1994},
\ben
\label{eq:def_Kpp}
  K_{pp}(R,\vec{x})= \frac{4}{3} \frac{1}{\delta (4-\delta^2) (4-\delta)} \frac{V_a^2 p^2}{K(R)}\,,
\een
whose strength is mediated by  the speed of plasma waves $V_a$ (the Alfv\'enic speed)\footnote{The reacceleration is pinched in the Galactic plane, and therefore $V_a$ values in this model should be scaled by a factor $\sqrt{h/L}$ before any comparison against theoretical or observational constraints \cite{ThornburyEtAl2014,DruryEtAl2017}.}.

Based on the analysis of \ams{} B/C data \cite{GenoliniEtAl2019} (in a 1D model), three different transport schemes were introduced (\BIG{}, \SLIM{}, and \QUAINT{}), with the presence or absence of a low-energy break or reacceleration. The fixed and free parameters of these configurations are reported in Table~\ref{tab:free_params}, and they are also used for the 2D model here (see next section).
\begin{table}
\begin{center}
\caption{Fixed (numbers) and free ($\checkmark$) parameters in Eq.~(\ref{eq:def_K}) for the transport configurations  \BIG{}, \SLIM{}, and \QUAINT{} \cite{GenoliniEtAl2019}. See Sect.~\ref{sapp:covmat} for the best-fit values of the free parameters.}
\label{tab:free_params}
\begin{tabular}{clccc}
\hline\hline
Parameters   & {\hspace{0.1cm}} Units & \BIG{}    & \SLIM{}     & \QUAINT{}  \\
\hline
 $\eta$     &      & 1          &  1         & \checkmark \\
 $\delta_{\rm l}$ & & \checkmark & \checkmark & n/a        \\
 $s_{\rm l}$      & & 0.05       &  0.05      & n/a        \\
 $R_{\rm l}$ & {\hspace{0.1cm}} {\scriptsize [GV]}      & \checkmark & \checkmark & n/a\footnote{In practice for \QUAINT~the rigidity $R_{\rm l}$ is set to 0 in Eq.~(\ref{eq:def_K}).}        \\
$V_{a}$  & {\hspace{0.1cm}} {\scriptsize [km/s]}    & \checkmark & n/a        & \checkmark \\
$V_{\rm c}$ & {\hspace{0.1cm}} {\scriptsize [km/s]}      & \checkmark & n/a        & \checkmark \\
 $K_0$   &     {\hspace{0.1cm}} {\scriptsize [kpc$^2$/Myr]}     & \checkmark & \checkmark & \checkmark \\
 $\delta$    &     & \checkmark & \checkmark & \checkmark \\
$\Delta_{\rm h}$  & & $0.18$      & $0.19$    & $0.17$     \\
$R_{\rm h}$ &  {\hspace{0.1cm}} {\scriptsize [GV]}  & $247$       & $237$     & $270 $     \\
$s_{\rm h}$   &    & $0.04$      & $0.04$    & $ 0.04$    \\
$L$      &  {\hspace{0.1cm}}  {\scriptsize [kpc]}     & \checkmark & \checkmark & \checkmark \\
\hline
\end{tabular}
\end{center}
\end{table}

\subsection{Best-fit parameters from AMS-02 LiBeB data}
\label{sapp:1Dto2D}

A state-of-the-art methodology to analyze \ams{} data and constrain propagation parameters---accounting for uncertainties in production cross sections and covariance matrix of CR data uncertainties---was proposed in \cite{DeromeEtAl2019}. This methodology was used in \cite{WeinrichEtAl2020,WeinrichEtAl2020b} for the combined analysis of \ams{} Li/C, Be/B, and B/C data, to obtain the transport parameters in several configurations (recalled in Table~\ref{tab:free_params}). However, this analysis was performed in a 1D model, whereas for dark matter studies, a 2D version of the model is mandatory. Indeed, in the 1D model the galactic diffusive halo is considered as an infinite slab in the radial direction, and the vertical direction $z$ is the only variable. As the DM distribution has a non-trivial radial profile, an appropriate modeling in the $r$ direction becomes necessary.

Compared to the 1D model, the 2D model requires two extra parameters: a radial boundary (taken at $R=20$~kpc) and a radial distribution of CR sources. The latter, for the case of astrophysical sources, can be estimated from the distribution of supernova remnants \cite{CaseEtAl1998,Green2015} or pulsars \cite{Yusifov2004,LorimerEtAl2006}. In order to decide which transport parameters to use for this analysis, we performed a comparison (with the \usine{} package \cite{Maurin2020}) of the B/C predictions obtained in the 1D model and in the 2D model for various assumptions:
\begin{itemize}
   \item  Using a constant CR source distribution in the 2D model gives similar results as in the 1D model as long as $L\ll L_R$ (where $L_R=R-R_\odot$), i.e.~if the distance between the observer and the radial boundary is much smaller than the halo size. Indeed, contributions from sources farther away than a boundary are exponentially suppressed in diffusion processes \cite{TailletEtAl2003}. If $L_R\sim L$, the radial boundary becomes a new suppression scale, which breaks down the equivalence with the 1D model. We observe a few percent (energy dependent) impact on the B/C calculation for $L=5$~kpc. This amplitude of the effect is correlated with the value of $L$.
   \item Using a more realistic spatial distribution of sources, the above radial boundary effect is mitigated (using \cite{Green2015}) or slightly amplified (using \cite{Yusifov2004,LorimerEtAl2006}). This is understood as the sharply decreasing source distribution with the modeling radius implies that fewer sources are suppressed by the radial boundary. A detailed analysis of these subtle effects goes beyond the scope of this paper and will be discussed elsewhere.
\end{itemize}

To summarize, the 1D model or the 2D version with a realistic source distribution are expected to provide similar results (at a few percent level), and thus similar best-fit transport parameters and uncertainties; we explicitly checked it for the \slim{} model using the radial distribution from \cite{Green2015} with $R=20$~kpc. For these reasons, we conclude that the 1D model parameters found in \cite{WeinrichEtAl2020,WeinrichEtAl2020b} can be used `as is' in the context of 2D models.

\subsection{Best-fit values and covariance matrix of uncertainties for \SLIM{}/\BIG{}/\QUAINT{}}
\label{sapp:covmat}

We provide below the best-fit parameter values of the model and their correlation matrix of uncertainties. Both come from the analyses discussed in \cite{WeinrichEtAl2020,WeinrichEtAl2020b}. We stress that all the values below correspond to the combined analysis using Li/B, Be/B, B/C \ams{} data, and low-energy $^{10}$Be/Be and $^{10}$Be/$^9$Be data ({\sc Ace-Cris, Ace-Sis, Imp7\&8, Isee3-Hkh, Ulysses-Het, Isomax}, and \voyager{} 1\&2, see details and references in \cite{WeinrichEtAl2020b}), in order to obtain the most stringent constraints on the halo size $L$.

Actually, in \cite{WeinrichEtAl2020}, only the best-fit transport parameters and uncertainties were given, whereas only the best-fit halo size constraint and its uncertainty were given in \cite{WeinrichEtAl2020b}, and for the analysis of different datasets. To ensure that the correct parameters are used, we gather them all in one place here. We provide in addition the full correlations matrix of uncertainties, to go beyond the \min-\med-\max{} benchmark models (see next section). How this matrix was obtained in the original analysis, and further checks on its validity are presented in Sect.~\ref{app:mcmc}. 

In the matrices shown below, the rows and columns correspond to the ordering of the best-fit parameters. We stress that parameters with a nearly Gaussian probability distribution function (see Sect.~\ref{app:mcmc}) in the covariance matrices are $\log_{10}[L/(1\,{\rm kpc})]$ and $\log_{10}[K_0/(1\,{\rm kpc}^2\,{\rm Myr}^{-1})]$---$\log_{10}L$ and $\log_{10}K_0$ for short---, not $L$ and $K_0$.

\

\paragraph*{Parameter values and covariance matrix for \SLIM{}}
\[
\small
\begin{matrix}
\log_{10} L & \delta & \log_{10} K_0 & R_{\rm l} & \delta_{\rm l}\\
0.668 & 0.499 & -1.444 & 4.482 & -1.110
\end{matrix}
\]
\[
\footnotesize
\begin{pmatrix}
+1.13\textrm{e-2} & -2.05\textrm{e-4} & +1.10\textrm{e-2} & +1.96\textrm{e-3} & +2.41\textrm{e-3}\\
-2.05\textrm{e-4} & +1.06\textrm{e-4} & -3.91\textrm{e-4} & +1.03\textrm{e-6} & -3.38\textrm{e-4}\\
+1.10\textrm{e-2} & -3.91\textrm{e-4} & +1.12\textrm{e-2} & +1.79\textrm{e-3} & +3.28\textrm{e-3}\\
+1.96\textrm{e-3} & +1.03\textrm{e-6} & +1.79\textrm{e-3} & +2.80\textrm{e-2} & +1.42\textrm{e-2}\\
+2.41\textrm{e-3} & -3.38\textrm{e-4} & +3.28\textrm{e-3} & +1.42\textrm{e-2} & +1.88\textrm{e-2}
\end{pmatrix}
\]

\

\paragraph*{Parameter values and covariance matrix for \BIG{}}
\[
\small
\begin{matrix}
\log_{10} L & \delta & \log_{10} K_0 & V_\mathrm{A} & R_{\rm l} & \delta_{\rm l} & V_\mathrm{c}\\
0.667 & 0.498 & -1.446 & 5.000 & 4.493 & -1.102 & 0.140
\end{matrix}
\]
\[
\scriptsize
\begin{pmatrix}
+4.20\textrm{e-3}\!\!&\!\!+3.53\textrm{e-4}\!\!&\!\!+3.94\textrm{e-3}\!\!&\!\!+1.49\textrm{e-2}\!\!&\!\!+2.57\textrm{e-3}\!\!&\!\!-1.48\textrm{e-3}\!\!&\!\!+4.56\textrm{e-2}\\
+3.53\textrm{e-4}\!\!&\!\!+4.19\textrm{e-4}\!\!&\!\!+7.06\textrm{e-4}\!\!&\!\!+3.96\textrm{e-3}\!\!&\!\!+2.89\textrm{e-3}\!\!&\!\!-1.41\textrm{e-3}\!\!&\!\!+3.24\textrm{e-3}\\
+3.94\textrm{e-3}\!\!&\!\!+7.06\textrm{e-4}\!\!&\!\!+5.48\textrm{e-3}\!\!&\!\!+1.86\textrm{e-2}\!\!&\!\!+4.67\textrm{e-3}\!\!&\!\!-2.18\textrm{e-3}\!\!&\!\!+4.57\textrm{e-2}\\
+1.49\textrm{e-2}\!\!&\!\!+3.96\textrm{e-3}\!\!&\!\!+1.86\textrm{e-2}\!\!&\!\!+2.02\textrm{e+1}\!\!&\!\!+6.00\textrm{e-2}\!\!&\!\!-4.52\textrm{e-2}\!\!&\!\!+1.96\textrm{e-1}\\
+2.57\textrm{e-3}\!\!&\!\!+2.89\textrm{e-3}\!\!&\!\!+4.67\textrm{e-3}\!\!&\!\!+6.00\textrm{e-2}\!\!&\!\!+2.92\textrm{e-2}\!\!&\!\!-1.30\textrm{e-2}\!\!&\!\!+2.15\textrm{e-2}\\
-1.48\textrm{e-3}\!\!&\!\!-1.41\textrm{e-3}\!\!&\!\!-2.18\textrm{e-3}\!\!&\!\!-4.52\textrm{e-2}\!\!&\!\!-1.30\textrm{e-2}\!\!&\!\!+2.11\textrm{e-2}\!\!&\!\!-1.28\textrm{e-2}\\
+4.56\textrm{e-2}\!\!&\!\!+3.24\textrm{e-3}\!\!&\!\!+4.57\textrm{e-2}\!\!&\!\!+1.96\textrm{e-1}\!\!&\!\!+2.15\textrm{e-2}\!\!&\!\!-1.28\textrm{e-2}\!\!&\!\!+1.86\textrm{e+0}
\end{pmatrix}
\]

\

\paragraph*{Parameter values and covariance matrix  for \QUAINT{}}
\[
\small
\begin{matrix}
\log_{10} L & \delta & \log_{10} K_0 & V_\mathrm{A} & V_\mathrm{c} & \eta_t\\
0.611 & 0.458 & -1.405 & 52.208 & 0.000 & -1.945
\end{matrix}
\]
\[
\footnotesize
\begin{pmatrix}
+6.12\textrm{e-3} & +7.07\textrm{e-4} & +4.75\textrm{e-3} & +3.12\textrm{e-1} & +2.60\textrm{e-3} & +1.82\textrm{e-2}\\
+7.07\textrm{e-4} & +6.43\textrm{e-4} & +1.26\textrm{e-3} & +2.87\textrm{e-1} & +8.72\textrm{e-4} & +1.45\textrm{e-2}\\
+4.75\textrm{e-3} & +1.26\textrm{e-3} & +7.58\textrm{e-3} & +9.22\textrm{e-1} & +1.76\textrm{e-3} & +3.06\textrm{e-2}\\
+3.12\textrm{e-1} & +2.87\textrm{e-1} & +9.22\textrm{e-1} & +1.95\textrm{e+2} & +7.14\textrm{e-1} & +7.47\textrm{e+0}\\
+2.60\textrm{e-3} & +8.72\textrm{e-4} & +1.76\textrm{e-3} & +7.14\textrm{e-1} & +3.31\textrm{e-1} & +3.01\textrm{e-2}\\
+1.82\textrm{e-2} & +1.45\textrm{e-2} & +3.06\textrm{e-2} & +7.47\textrm{e+0} & +3.01\textrm{e-2} & +5.02\textrm{e-1}
\end{pmatrix}
\]

\subsection{Drawing from the covariance matrix in practice}
\label{sapp:draw_params}
For a DM analysis using the full statistical information on the transport parameters, one needs to draw from best-fit values and the associated covariance matrix of uncertainties presented in \citeapp{sapp:covmat}. The following {\tt numpy} \cite{Harris2020} command can be used for instance:
\[
   {\tt random.multivariate\_normal(pars, cov, size=N)}, 
\]
where {\tt pars} is an array of the best-fit parameters, {\tt cov} is the associated covariance matrix, and {\tt N} is the number of samples to draw.

By construction, the parameter distributions are symmetric, so that non-physical negative values can be obtained for $V_a$ and $V_c$: the sample behaving so should be discarded (or alternatively used with $V_a$ and $V_c$ set to zero). There are also a few other important points to be kept in mind:
\begin{itemize}
   \item Full list of parameters: for a full description of the model, and in particular of the diffusion coefficient Eq.~(\ref{eq:def_K}), the parameters drawn must be complemented by the `fixed' parameter values given in Table~\ref{tab:free_params} (no associated covariance matrix of uncertainties);
   \item Meaning of $V_a<5$~km~s$^{-1}$: as discussed in \cite{DeromeEtAl2019}, we enforce (obviously positive but also) non-null values of $V_a$ for numerical issues. As a result, any value of $V_a$ smaller than 5~km~s$^{-1}$ should be understood as $V_a = 0$.
   \item Specific form of $K(R)$ for \QUAINT{}: this model does not enable a low energy break, and one needs to remove in Eq.~(\ref{eq:def_K}) the associated terms (first square bracket), or alternatively, to set $R_l$ to zero.
\end{itemize}

\subsection{Parametrisation for reference secondary $\bar{p}$ and $e^+$}
\label{sapp:pbar_formula}

In order to set constraints on DM candidates, it is mandatory to know the astrophysical secondary fluxes of CR on top of which the DM signal is searched for. To enable this kind of searches, we provide here a parametric formula for such secondary backgrounds:
\begin{equation}
 \log_{10}\!\!\left[\frac{\phi^{\rm IS}}{1/({\rm U_x~m}^{2}{\rm s~sr})}\right] \!\!= c_0 \!+\! \sum_{i=1}^{10} \! c_i \! \left[\log_{10}\!\!\left(\frac{x}{x_{\rm th}}\right)\right]^i,
\label{eq:fit_formula}
\end{equation}
where $x$ (and the corresponding unit $U_x$) is the rigidity $R$ ($U_x=$~GV) when considering $\bar{p}$, or the kinetic energy $E_k$ ($U_x=$~GeV) when considering positrons.

\paragraph{Coefficients for $\bar{p}$.} 
The fit is based on the calculation presented in \cite{BoudaudEtAl2020}. The coefficients to apply below and above $x_{\rm th}=8$~GV are given in Table~\ref{tab:pbar_coeffs}: eleven coefficients were needed to reproduce the fluxes with a precision better than $1\%$, and the formula applies to IS rigidities from 0.9~GV to 10~TV.

\begingroup
\footnotesize
\begin{table}[t]
\caption{Coefficients for the reference secondary $\bar{p}$ IS fluxes of \cite{BoudaudEtAl2020} in (GV~m$^{2}$~s~sr)$^{-1}$. This parametrisation, see Eq.~(\ref{eq:fit_formula}), is valid for rigidities from 0.9~GV to 10~TV. These coefficients hold for \min, \med, and \max\ since the differences are small. \label{tab:pbar_coeffs}}
\begin{tabular}{ll@{\hskip 1cm}ll}
\hline\hline
\multicolumn{2}{c}{($< 8$ GV)} & \multicolumn{2}{c}{($\ge8$ GV)}\\
\hline
\multicolumn{4}{c}{\SLIM{}}\\
$c_{0}$  & $-2.059841$  & $c_{0}$  & $-2.041091$\\
$c_{1}$  & $-3.742100$  & $c_{1}$  & $-2.125402$\\
$c_{2}$  & $-3.853337\times10^{1}$  & $c_{2}$  & $-1.754931$\\
$c_{3}$  & $-3.921245\times10^{2}$  & $c_{3}$  & $+2.370157$\\
$c_{4}$  & $-2.139289\times10^{3}$  & $c_{4}$  & $-2.168217$\\
$c_{5}$  & $-6.615346\times10^{3}$  & $c_{5}$  & $+1.164920$\\
$c_{6}$  & $-1.245434\times10^{4}$  & $c_{6}$  & $-2.432764\times10^{-1}$\\
$c_{7}$  & $-1.457810\times10^{4}$  & $c_{7}$  & $-6.773555\times10^{-2}$\\
$c_{8}$  & $-1.037382\times10^{4}$  & $c_{8}$  & $+5.097495\times10^{-2}$\\
$c_{9}$  & $-4.114799\times10^{3}$  & $c_{9}$  & $-1.102931\times10^{-2}$\\
$c_{10}$ & $-6.986690\times10^{2}$  & $c_{10}$ & $+8.537912\times10^{-4}$\\[3mm]
\hline\hline
\end{tabular}
\end{table}
\endgroup

In principle, for an exotic flux calculation from a given set of transport parameters (drawn from the covariance matrix above), the secondary flux should be re-calculated. However, the secondary flux calculation requires the full propagation of all nuclear species and the inclusion of many extra ingredients. Because this complicates and greatly slows down the calculation (compared to the exotic-flux-only calculation), and because the secondary flux `only' varies within $10-20\%$ over the transport parameter space \cite{BoudaudEtAl2020}, the use of the reference secondary flux formula remains useful and a very good first approximation to quickly explore the parameters space of new physics models.

\paragraph{Coefficients for $e^+$}
The fit is based on the calculation we presented in \cite{WeinrichEtAl2020b}. The coefficients to apply below and above $x_{\rm th}=1$~GeV, are presented in Table~\ref{tab:posit_coeffs}: we also took eleven coefficients to reproduce the fluxes at the percent level precision, and the formula applies to IS kinetic energies from 2~MeV to 1~TeV.

\begingroup
\footnotesize
\begin{table}[t]
\caption{Coefficients for the reference secondary $e^+$ IS fluxes of \cite{WeinrichEtAl2020b} in (GeV~m$^{2}$~s~sr)$^{-1}$. This parametrisation, see Eq.~(\ref{eq:fit_formula}), is valid for kinetic energies from 0.002~GeV to 1~TeV. \label{tab:posit_coeffs}}
\begin{tabular}{ll@{\hskip 1cm}ll}
\hline\hline
\multicolumn{2}{c}{($< 1$ GeV)} & \multicolumn{2}{c}{($\ge1$ GeV)}\\
\hline
\multicolumn{4}{c}{\SLIM{} (\min{})}\\
$c_{0}$  & $+4.996550\times10^{-1}$  & $c_{0}$  & $+5.075677\times10^{-1}$\\
$c_{1}$  & $-1.128052$  & $c_{1}$  & $-1.443865$\\
$c_{2}$  & $+8.821920\times10^{-1}$  & $c_{2}$  & $+1.877049$\\
$c_{3}$  & $+1.157137\times10^{1}$  & $c_{3}$  & $-1.066219\times10^{1}$\\
$c_{4}$  & $+3.806619\times10^{1}$  & $c_{4}$  & $+1.319804\times10^{1}$\\
$c_{5}$  & $+6.252870\times10^{1}$  & $c_{5}$  & $-6.512217$\\
$c_{6}$  & $+5.736342\times10^{1}$  & $c_{6}$  & $-8.868464\times10^{-3}$\\
$c_{7}$  & $+3.097409\times10^{1}$  & $c_{7}$  & $+1.500031$\\
$c_{8}$  & $+9.799701$  & $c_{8}$  & $-7.040355\times10^{-1}$\\
$c_{9}$  & $+1.682151$  & $c_{9}$  & $+1.401731\times10^{-1}$\\
$c_{10}$ & $+1.210159\times10^{-1}$  & $c_{10}$ & $-1.074125\times10^{-2}$\\[3mm]
\multicolumn{4}{c}{\SLIM{} (\med{})}\\
$c_{0}$  & $+6.525177\times10^{-1}$  & $c_{0}$  & $+6.461406\times10^{-1}$\\
$c_{1}$  & $-1.588678$  & $c_{1}$  & $-1.384556$\\
$c_{2}$  & $+4.078335\times10^{-1}$  & $c_{2}$  & $-4.774200$\\
$c_{3}$  & $+1.069849\times10^{1}$  & $c_{3}$  & $+1.366313\times10^{1}$\\
$c_{4}$  & $+3.723089\times10^{1}$  & $c_{4}$  & $-2.902370\times10^{1}$\\
$c_{5}$  & $+6.209276\times10^{1}$  & $c_{5}$  & $+3.638689\times10^{1}$\\
$c_{6}$  & $+5.723649\times10^{1}$  & $c_{6}$  & $-2.734189\times10^{1}$\\
$c_{7}$  & $+3.094710\times10^{1}$  & $c_{7}$  & $+1.258343\times10^{1}$\\
$c_{8}$  & $+9.790895$  & $c_{8}$  & $-3.485906$\\
$c_{9}$  & $+1.679555$  & $c_{9}$  & $+5.346104\times10^{-1}$\\
$c_{10}$ & $+1.207142\times10^{-1}$  & $c_{10}$ & $-3.491387\times10^{-2}$\\[3mm]
\multicolumn{4}{c}{\SLIM{} (\max{})}\\
$c_{0}$  & $+7.593070\times10^{-1}$  & $c_{0}$  & $+7.480398\times10^{-1}$\\
$c_{1}$  & $-2.339091$  & $c_{1}$  & $-1.915233$\\
$c_{2}$  & $+9.758912\times10^{-6}$  & $c_{2}$  & $-6.642892$\\
$c_{3}$  & $+1.276888\times10^{1}$  & $c_{3}$  & $+2.428124\times10^{1}$\\
$c_{4}$  & $+4.395641\times10^{1}$  & $c_{4}$  & $-5.077192\times10^{1}$\\
$c_{5}$  & $+7.102167\times10^{1}$  & $c_{5}$  & $+6.136908\times10^{1}$\\
$c_{6}$  & $+6.382395\times10^{1}$  & $c_{6}$  & $-4.511720\times10^{1}$\\
$c_{7}$  & $+3.384382\times10^{1}$  & $c_{7}$  & $+2.058580\times10^{1}$\\
$c_{8}$  & $+1.054571\times10^{1}$  & $c_{8}$  & $-5.705532$\\
$c_{9}$  & $+1.787170$  & $c_{9}$  & $+8.807590\times10^{-1}$\\
$c_{10}$ & $+1.271850\times10^{-1}$  & $c_{10}$ & $-5.812954\times10^{-2}$\\[3mm]
\hline\hline
\end{tabular}
\end{table}
\endgroup

We stress that this positron flux only accounts for the astrophysical secondary flux, but does not account for astrophysical primary contributions \cite{WeinrichEtAl2020b}.\\

Note that these parametrisations concern only the secondary predictions for the \SLIM~model. For completeness we also provide on reasonable request, the secondary TOA fluxes for $e^+$ and $\bar{p}$ for the two other models (\BIG, \QUAINT) in the form of ancillary files. 

\begin{figure*}[t!]
  \includegraphics[width=0.85\textwidth]{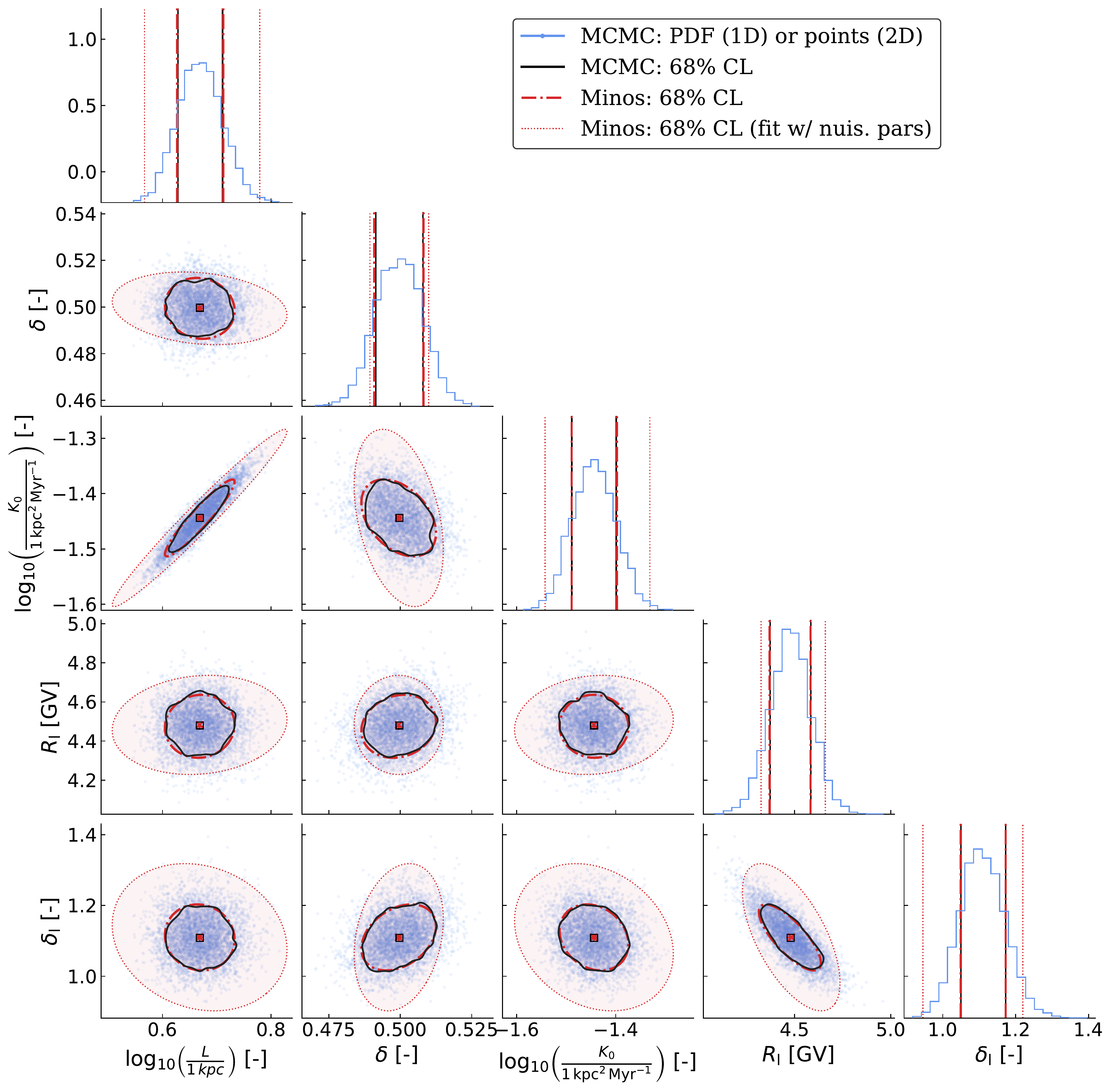}
  \caption{Comparison of constraints on transport parameters obtained from an \mcmc{} engine or via the \hesse{}-\minos{} algorithms. In off-diagonal plots, the black squares and red crosses show the best-fit values from the former and latter results respectively, \mcmc{} points are shown in blue (from which 68\% confidence level contours are extracted in black), and red ellipses are based on symmetrized \minos{} errors (dash-dotted or dotted lines for the fit accounting for nuisance parameters). In diagonal plots,  the blue histograms corresponds to the 1D probability distribution function (\mcmc{}), and the 68\% CL contours are shown as vertical blue lines (\mcmc{}) and red vertical lines (\hesse{}-\minos{}). We recall that the 1D 68\% CL on any given parameter is not expected to match the projected value obtained from the 2D 68\% CL. See text for discussion.}
  \label{fig:app_MCMC}
\end{figure*}

\section{MCMC versus Minos results}
\label{app:mcmc}

In this appendix, we show that the covariance matrix of uncertainties reconstructed with the help of \minuit~\cite{JamesEtAl1975} provides a sufficient description of the behavior of the transport parameter uncertainties.

To do so, we perform a Markov Chain Monte Carlo (\mcmc) analysis of the transport parameters and compare the results with those obtained from the \hesse\ and \minos\ algorithms in \minuit{}. In the latter approach, \hesse\ provides a covariance (symmetric) matrix of uncertainties on the model parameters, but whose uncertainties (from \migrad{}) are not very robust. In order to obtain a more reliable covariance matrix, we rescale the \hesse\ covariance matrix by the symmetrized \minos\ errors; doing so ensures that we keep the `correct' correlations, but now with our best knowledge on the error size. The \mcmc\ engine provides the full PDF on the parameters. If the PDFs are Gaussian, we should obtain similar confidence levels and contours on the parameters from both approaches.

This analysis follows closely that of \cite{WeinrichEtAl2020b} and we only briefly highlight the most important elements or differences. The determination of the transport parameters and halo size of the Galaxy ($L$) are based on \ams{} Li/B, Be/B, and B/C data \cite{AguilarEtAl2018}, and the analysis with \minuit\ matches exactly that described in \cite{WeinrichEtAl2020b}. For the \mcmc\ analysis, we relied on the \pymc\ package\footnote{\url{https://docs.pymc.io/}} and its Metropolis-Hastings sampler\footnote{Technically, we took advantage of \pybind, \url{https://pybind11.readthedocs.io}, to enable the interface with python libraries of the C++ propagation code \usine~\cite{Maurin2020}.}. More details and references on this algorithm and the various steps associated with the post-processing of the chain (burn-in length, trimming, etc.) can be found for instance in \cite{PutzeEtAl2009}, where a similar algorithm and approach was first considered (in a cosmic-ray propagation context).

We show in \citefig{fig:app_MCMC} the result of the comparison on the propagation configuration \SLIM\ only. In the latter, the free transport parameters are the normalization and slope of the diffusion coefficient at intermediate rigidities ($K_0$ and $\delta$), the position and strength of a possible low energy break ($R_{\rm l}$ and $\delta_{\rm l}$), and the halo size of the Galaxy ($L$). First, as a sanity check, the red crosses and black squares in 2D plots of the parameters (off-diagonal) show that both approaches provide the same best-fit values; we also recognize the tight correlation between $K_0$ and $L$ parameters.
On the same off-diagonal plots, the black solid lines correspond to $68\%$ confidence level (CL) contours of the \mcmc\ analysis, while the red dash-dotted lines correspond to $68\%$ CL ellipses from the \migrad-\minos\ approach, and both match very well. A similar information is shown in the 1D plots (diagonal), where the red dash-dotted (from \migrad-\minos) is superimposed on the black solid lines (\mcmc).
The full information on the PDF is provided by the blue histograms. As already observed in \cite{PutzeEtAl2010} (see their App.~C) and also seen here, the transport parameters are Gaussian at first order. Hence, using the covariance matrix of uncertainties (as built above) is a very good approximation to using the full PDF information, and it will eventually depart from the latter only if too large confidence levels are considered.

There are possibly a few caveats to these conclusions. Firstly, due to fact that \mcmc\ analyses are computationally demanding, we only performed the comparison for the \SLIM\ model. However, we do not expect different behaviors for the other configurations (\BIG\ and \QUAINT), especially for the most relevant parameters in a dark matter context ($K_0$, $\delta$, and $L$), as these parameters behave similarly in terms of their \minos{} uncertainties \cite{WeinrichEtAl2020b}. Secondly, the above comparisons were made without nuisance parameters (nuclear cross sections and Solar modulation parameters), despite their importance for the determination of the transport parameters \cite{DeromeEtAl2019,GenoliniEtAl2019,WeinrichEtAl2020}. This is illustrated comparing the red dash-dotted and dotted lines in \citefig{fig:app_MCMC}: accounting for nuisance parameters enlarges the contours (larger uncertainties) for almost all parameters. We tried to run \mcmc\ chains with nuisance parameters, but the latter strongly increase the correlations length of the chains (they are correlated with other parameters and partly degenerate). So far, we did not succeed in obtaining reliable results for an \mcmc\ analysis. Lastly, the comparisons were made in the context of the 1D diffusion model, and not in the 2D one that is used in the main text. However, as shown in \cite{PutzeEtAl2010} (with an \mcmc{} analysis), the propagation parameters are very similar whether determined in a 1D or 2D model, as long as $L\lesssim10$~kpc, which is the case here (see also the previous section).
\section{The {\min}, {\med}, and {\max} configurations for the {\QUAINT} and {\BIG} models}

\begin{figure}[t]
\includegraphics[width=0.87\columnwidth]{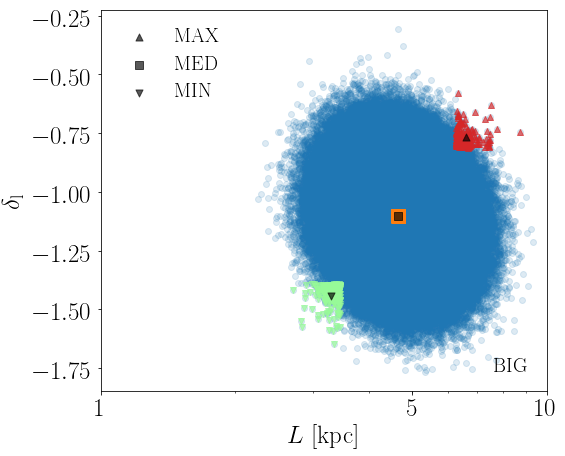}
\includegraphics[width=0.84\columnwidth]{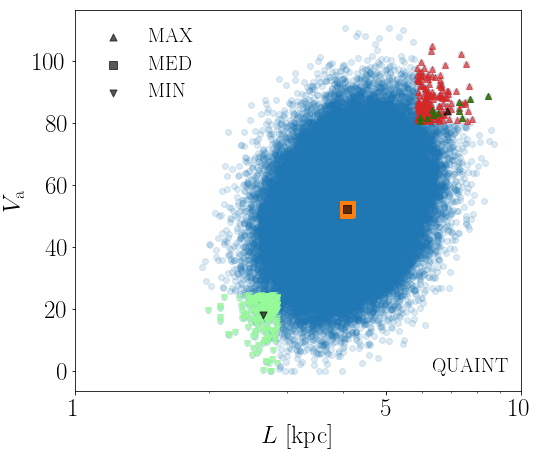}
\caption{
Same as in Fig.~\ref{fig:Def_MinMedMax_slim} for the {\big} (up) and {\quaint} (down) declensions of CR propagation models.
The constellations of blue dots contain each $10^{5}$ randomly drawn models.
The {\min}, {\med}, and {\max} sub-samples together with their barycentric configurations are defined as explained in Sec~\ref{sec:stats}. For {\big}, the selection is still built on the parameters $\log_{10}{L}$ and $\delta_{\rm l}$ whereas for {\quaint}, it is based on the couple $\log_{10}{L}$ and $V_{a}$. In the latter case, an additional skimming of the {\med} and {\max} populations is performed, requiring that the secondary positron flux does not exceed the observations~\cite{AguilarEtAl2019}. The configurations which actually pass this test are shown in dark-green.
}
\label{fig:Def_MinMedMax_big_quaint}
\end{figure}
\label{app:min_med_max_general}

The {\slim} and {\big} benchmarks are fairly similar. In the latter case, two additional parameters are introduced, i.e. the Alfv\'enic speed $V_{a}$ and the convective wind velocity $V_{c}$ to recover more easily the behavior of the B/C ratio in the GeV range. Actually, the low-energy parameters $\delta_{\rm l}$ and $R_{\rm l}$ are enough to reach a good agreement with data. That is why the values of $V_{a}$ and $V_{c}$ provided by the fits to CR nuclei are small, as showed in Table~\ref{tab:stat_parvalues_big}. In order to define the {\min}, {\med}, and {\max} models for the {\big} benchmark, we have proceeded as in the {\slim} case, using the parameters $\log_{10}{L}$ and $\delta_{\rm l}$. The result is showed in the top panel of Fig.~\ref{fig:Def_MinMedMax_big_quaint}. The values of the quantiles $q_{\textrm{\min{}}}$, $q_{\textrm{\med{}}}$ and $q_{\textrm{\max{}}}$ and of the width parameter $p$ are the same as those of Sec.~\ref{sec:stats}.

\begin{table}[t]
\caption{
Propagation parameters for the {\min}, {\med}, and {\max} configurations of the {\big} models.
}
\label{tab:stat_parvalues_big}
\begin{tabular}{p{0.8cm}p{0.9cm}p{0.9cm}p{1.3cm}p{0.9cm}p{0.9cm}p{0.9cm}p{0.8cm}}
\hline\hline
{\big} 	& $L$ 			& $\delta$ 	& $\log_{10} K_0$ 			& $V_{a}$ 	& $R_\mathrm{l}$ 	& $\delta_\mathrm{l}$ 	& $V_{\rm c}$ \\
	  	& {\scriptsize [kpc]} 	& 		 	& {\scriptsize [kpc$^2$/Myr]} 	& {\scriptsize [km/s]} 	& {\scriptsize [GV]}		&     &   {\scriptsize [km/s]}   \\
\hline
{\max} &     6.637     &     0.529     &     -1.286     &     6.002     &     4.755     &     -1.455     &     1.819 \\
{\med} &     4.645     &     0.498     &     -1.446     &     4.741     &     4.490     &     -1.102     &     0.459 \\
{\min} &     3.206     &     0.465     &     -1.616     &     4.277     &     4.208     &     -0.742     &     0.066 \\
\hline
\end{tabular}
\end{table}

The {\quaint} benchmark makes use of the low-energy parameters $V_{a}$, $V_{c}$ and $\eta$ and disregards $R_{\rm l}$ and $\delta_{\rm l}$. Reproducing the B/C GeV bump requires fairly large values of the Alfv\'enic speed $V_{a}$ as can be appreciated from Table~\ref{tab:stat_parvalues_quaint}. This parameter controls diffusive reacceleration which pushes sub-GeV CR species upward in the GeV energy region. We have used it together with $\log_{10}{L}$ to define the {\min}, {\med}, and {\max} sub-samples extracted from a population of $10^{5}$ randomly drawn {\quaint} models. The procedure is the same as before except that $\delta_{\rm l}$ has been replaced by $V_{a}$ as showed in the bottom panel of Fig.~\ref{fig:Def_MinMedMax_big_quaint}.

\begin{table}[t]
\caption{
Propagation parameters for the {\min}, {\med}, and {\max} configurations of the {\quaint} models.
}
\label{tab:stat_parvalues_quaint}
\begin{tabular}{p{1.2cm}p{0.9cm}p{0.9cm}p{1.3cm}p{1.1cm}p{0.9cm}p{1.0cm}}
\hline\hline
{\quaint} 	& $L$ 			& $\delta$ 	& $\log_{10} K_0$ 			& $V_{a}$ 	& $V_{\rm c}$ 	& $\eta$ 	\\
	  	& {\scriptsize [kpc]} 	& 		 	& {\scriptsize [kpc$^2$/Myr]} 	& {\scriptsize [km/s]} 	& {\scriptsize [km/s]}	&		\\
\hline
{\max} &     6.840     &     0.504     &     -1.092     &     83.929     &     0.469     &      -1.001 \\
{\med} &     4.080     &     0.451     &     -1.367     &     52.066     &     0.239     &      -2.156 \\
{\min} &     2.630     &     0.403     &     -1.643     &     18.389     &     0.151     &     -3.412 \\
\hline
\end{tabular}
\end{table}

\begin{figure*}[t]
\includegraphics[width=0.66\textwidth]{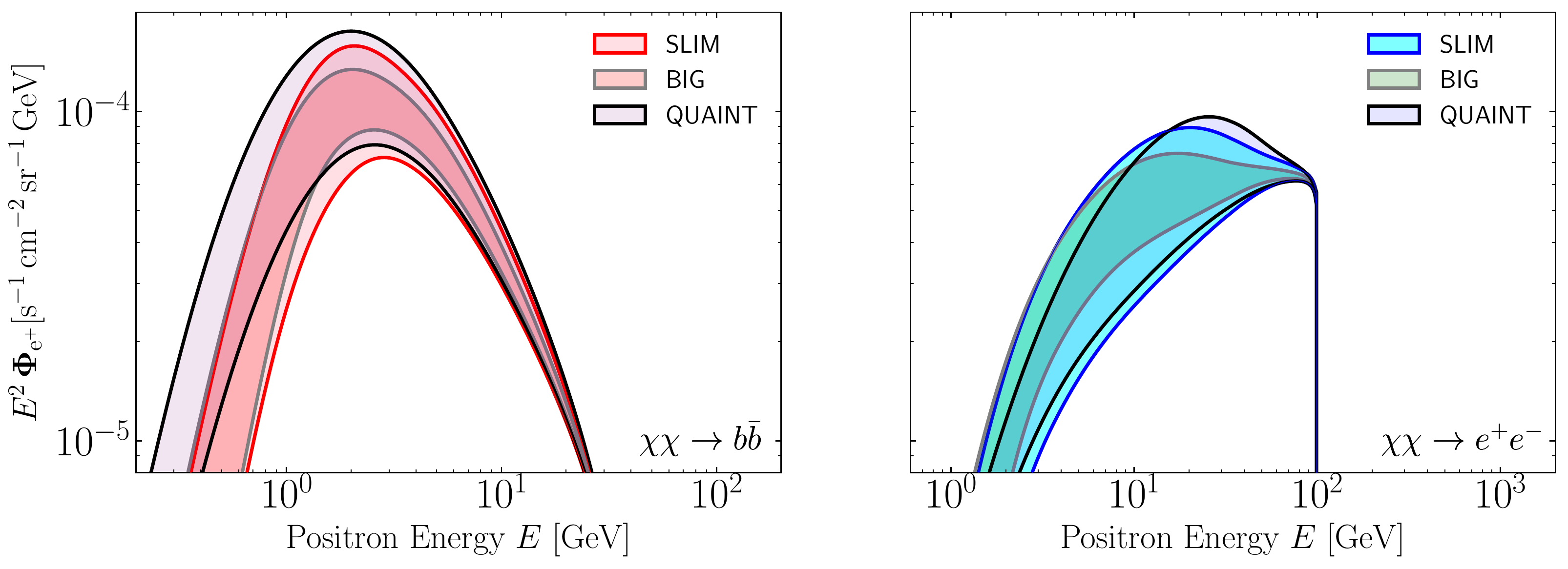}
\includegraphics[width=0.33\textwidth]{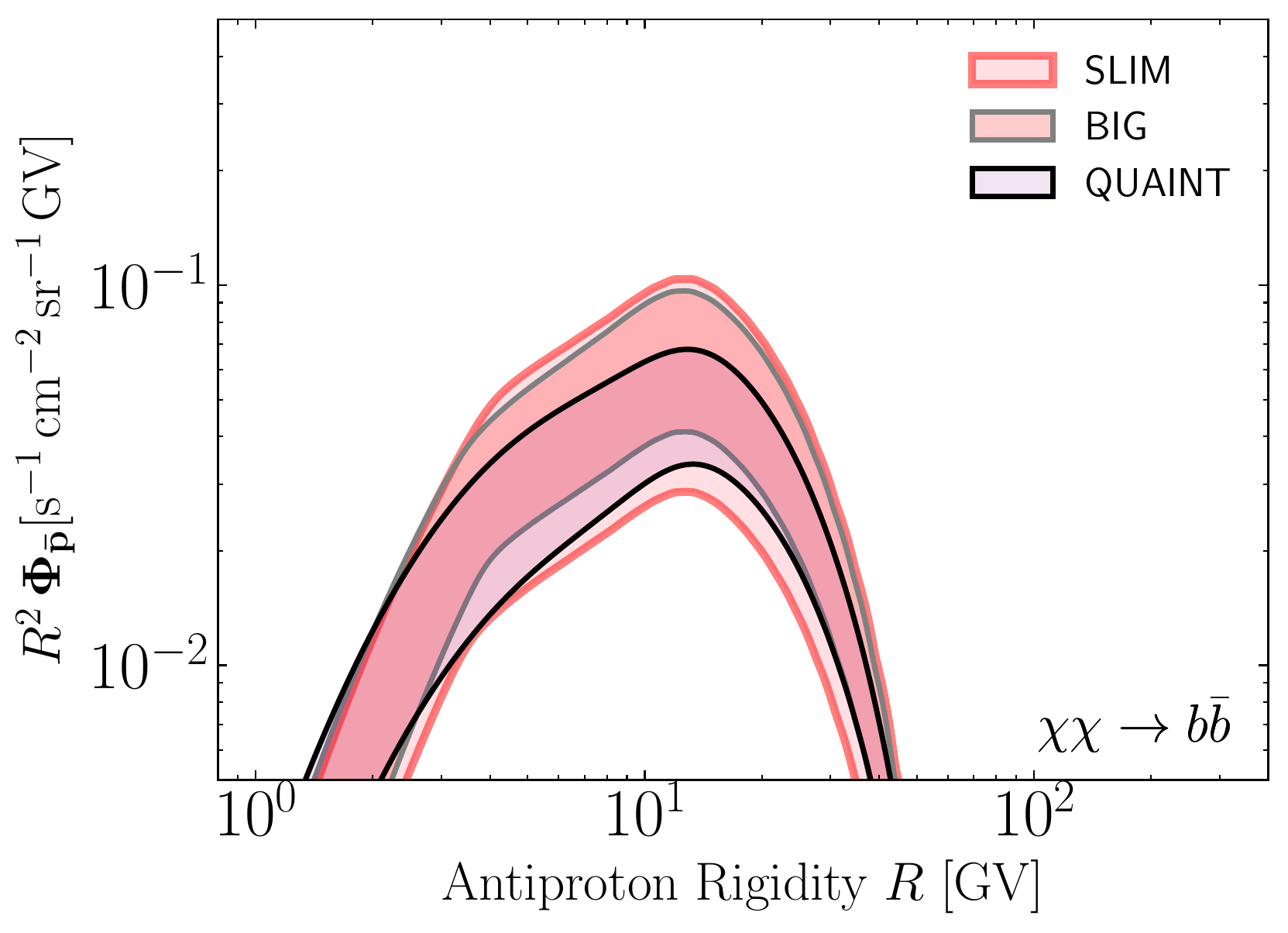}
\caption{
The theoretical uncertainty on primary fluxes owing to CR propagation for the three benchmarks {\slim}, {\big}, and {\quaint}. The panels are similar to those in Fig.~\ref{fig:comparisons_min_med_max_old_vs_new}. The pink and blue strips respectively correspond to $b \bar{b}$ and $e^{+}e^{-}$ channels. The right panel is devoted to antiprotons while the others feature results for positrons.
The bands nicely overlap each other. Although different in spirit, the three benchmarks yield similar predictions.
}
\label{fig:comparisons_min_med_max_SBQ}
\end{figure*}

There is however a slight complication that arises because the Alfv\'enic speed is high. For large values of $V_{a}$, the secondary positron flux exhibits, like the B/C ratio, a bump at a few GeV. In some cases, it even exceeds the observations. To remove these pathological models from the {\max} and {\med} sub-samples, where they tend to appear, we have required the secondary positron flux not to overshoot by more than 3 standard deviations the lowest \ams{} data point~\cite{AguilarEtAl2019}. To be conservative, we have used a Fisk potential $\Phi_F$ of $750\,$MV. The red and orange populations in the right panel of Fig.~\ref{fig:Def_MinMedMax_big_quaint} are the result of this skimming.

Finally, the theoretical uncertainties arising from CR propagation are summarized in Fig.~\ref{fig:comparisons_min_med_max_SBQ} for the three benchmarks {\slim}, {\big}, and {\quaint}. The pink and blue strips respectively stand for $b \bar{b}$ and $e^{+}e^{-}$ channels. Antiproton primary fluxes are presented in the right panel while the left and middle ones are devoted to positrons. This plot summarizes our entire analysis. The various bands overlap each other, indicating that in spite of their differences, the three benchmarks supply similar predictions for primary fluxes.

\newpage

\bibliographystyle{apsrev4-1}
\bibliography{minmedmax}

\end{document}